\documentclass[epj,nopacs]{svjour}
\usepackage{graphics}
\begin{document}
\title{Performance of the MALTA Telescope}

\author{Milou van Rijnbach\inst{1},\inst{2} \and Giuliano Gustavino\inst{1} \and Phil Allport\inst{3} \and Igancio Asensi\inst{1} \and Dumitru Vlad Berlea\inst{4} \and Daniela Bortoletto\inst{5} \and Craig Buttar\inst{6} \and Edoardo Charbon\inst{7} \and Florian Dachs\inst{1} \and Valerio Dao\inst{1} \and Dominik Dobrijevic\inst{1},\inst{8} \and Leyre Flores Sanz de Acedo\inst{1} \and Andrea Gabrielli\inst{1} \and Martin Gazi\inst{5} \and Laura Gonella\inst{3} \and Vicente Gonzalez\inst{9} \and Stefan Guidon\inst{1} \and Matt LeBlanc\inst{1} \and Heinz Pernegger\inst{1} \and Francesco Piro\inst{1}\inst{7} \and Petra Riedler\inst{1} \and Heidi Sandaker\inst{2} \and Abhishek Sharma\inst{1} \and Carlos Solans Sanchez\inst{1} \and Walter Snoeys\inst{1} \and Tomislav Suligoj\inst{8} \and Marcos Vazquez Nunez\inst{1},\inst{9} \and Julian Weick\inst{1},\inst{10} \and Steven Worm\inst{4} \and Abdelhak M. Zoubir\inst{10}.
}
\mail{milou.van.rijnbach@cern.ch}
\institute{CERN, Geneva, Switzerland \and University of Oslo, Oslo, Norway \and University of Birmingham, Birmingham, United Kingdom \and DESY, Zeuthen, Germany \and University of Oxford, Oxford, United Kingdom \and University of Glasgow, Glasgow, United Kingdom \and EPFL, Lausanne, Switzerland \and University of Zagreb, Zagreb, Croatia \and Universitat de València, València, Spain \and Technische Universität Darmstadt, Darmstadt, Germany}
\date{Received: date / Revised version: date}
%
\abstract{
MALTA is part of the Depleted Monolithic Active Pixel sensors designed in Tower 180nm CMOS imaging technology. A custom telescope with six MALTA planes has been developed for test beam campaigns at SPS, CERN, with the ability to host several devices under test. The telescope system has a dedicated custom readout, online monitoring integrated into DAQ with realtime hit map, time distribution and event hit multiplicity. It hosts a dedicated fully configurable trigger system enabling to trigger on coincidence between telescope planes and timing reference from a scintillator. The excellent time resolution performance allows for fast track reconstruction, due to the possibility to retain a low hit multiplicity per event which reduces the combinatorics. This paper reviews the architecture of the system and its performance during the 2021 and 2022 test beam campaign at the SPS North Area.%
} 
\maketitle

\section{Introduction}
\label{intro}

Beam telescopes are tracking detector systems used for the characterization of pixel detector prototypes for a wide range of applications. They allow for studies that are beyond feasible in a laboratory, by analysing the sensor's response to ionising particles, mimicking the operation conditions in real tracking detectors. Particles passing through the telescope are reconstructed and used in the characterisation of a given Device Under Test (DUT). \\ 

\begin{figure*}
\centering
\resizebox{0.75\textwidth}{!}{%
  \includegraphics{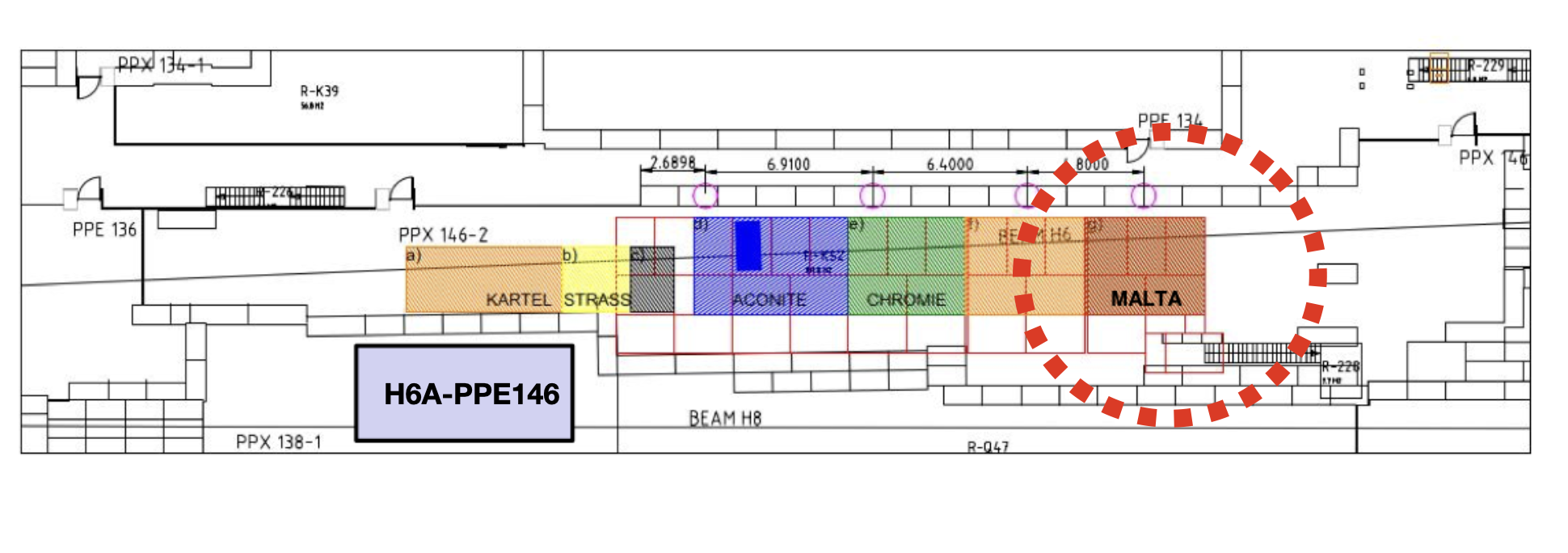}
}
\caption{A schematic image of the H6A/PPE146 beam line at the Super Proton Synchroton (SPS) at CERN. The experimental areas, including the permanent installation of the MALTA telescope downstream of the beam line, are indicated}
\label{beamline}       
\end{figure*}

\noindent MALTA is a Depleted Monolithic Active Pixel Sensor (DMPAS) designed in Tower 180 nm CMOS imaging technology. A custom telescope has been developed for test beam campaigns at the Super Proton Synchroton (SPS) at CERN using up to six MALTA tracking planes, a scintillator for precise timing reference, and the ability to host several DUTs. The telescope system has a dedicated readout, a fully configurable trigger system, an excellent time resolution, and a low hit multiplicity per event that reduces the combinatorics, which in turn allows for fast track reconstruction. Furthermore, there is the possibility for in-chip Region Of Interest (ROI) implementation, serving both large and small prototypes. The MALTA telescope is permanently installed at the H6 beamline (see Figure \ref{beamline}) in the North Area (NA), one of the secondary beam areas of SPS. The high-energy and high-resolution H6 beam line can transport mixed hadron beams within the range of 10-205 GeV/c to the respective experiments, with a beam intensity ranging between 2$\times$10$^5$  and 4.5$\times$10$^6$ particles per spill.  \\

\noindent This paper is dedicated to reviewing the architecture, operation, and performance of the MALTA telescope. First the MALTA sensor will be introduced, which will elaborate on the motivation for a MALTA based telescope and specifications on the tracking planes. Second, the main components of the telescope will be addressed, including the mechanical structure, trigger logic, and the data acquisition. Furthermore, the reconstruction and offline analysis framework will be discussed including the results on the spatial telescope resolution, followed by discussion on the timing resolution of the MALTA telescope. Finally, examples of DUT integration will be provided, focusing on results obtained with the MALTA2 sensor during the 2021 and 2022 test beam campaign and integration of various ATLAS R\&D prototypes within the custom Trigger Logic Unit (TLU).

\section{The MALTA Sensor}

MALTA is a DMAPS with small collection electrode, fabricated in Tower 180 nm CMOS imaging technology. The sensor was originally developed for its use in the Phase-II upgrade of ATLAS for the High Luminosity LHC \cite{atlas2017technical} and for other future collider experiments, the latter being still relevant for possible applications of MALTA. The large interest in monolithic pixel sensors has been driven by the possibility to minimize the material budget, reduce the production effort, and lowering the costs as the readout electronics and sensor are fabricated in the same silicon wafer. More specifically, the MALTA pixel sensor offers excellent radiation hardness up to about 100 Mrad Total Ionizing Dose (TID) and greater than 1$\times$10$^{15}$ 1 MeV n$_{eq}/{cm}^2$ in Non-Ionizing Energy Loss (NIEL) with a fast charge collection \cite{pernegger2017first,snoeys2017process}, as will be explained further below. \\

\noindent The MALTA matrix consists of 512$\times$512 pixels with a pixel pitch of 36.4 $\mu$m. The size of the small, octagonal-shaped collection electrode (diameter of 2 $\mu$m) results in a small capacitance, which consequently minimizes noise and allows for low power dissipation (10~mW/cm$^2$ digital and 70 mW/cm$^2$ analog power). The asynchronous readout transmits the hit information directly from chip to periphery and consequently through 37 parallel output signals (2 ns output signal length). The asynchronous readout avoids the distribution of a high frequency clock signals across the matrix, which minimizes analog-digital cross-talk. Pixels are organised in groups of 2$\times$8 and hits from a pixel are sent to a reference pulse generator which is common within the group. A reference pulse is generated, which is appended to the pixel and group address, respectively 16-bit and 5-bit. The hits are distributed in two parallel 22-bit wide busses, one for even groups and the other for odd groups \cite{cardella2019malta}. The benefit of this distinction lies in the fact that adjacent groups cannot share the same bus, reducing cross-talk on the hit address bus. Due to the area constraints of the chip, no Time-over-Threshold (ToT) information is directly available from the chip.  \\

\noindent Several studies \cite{cardella2019malta,caicedo2019monopix,schioppa2020measurement} have reported on the performance of irradiated DMAPS with small collection electrode, specifically on the sensors with the standard modified process modification (STD), where a low dose n-type blanket implant is introduced across the full pixel matrix. The studies showed that the detection efficiency for these type of sensors is compromised in the pixel corners. Therefore, the pixel design and sensor processing has been modified such that these deficiencies could be overcome while maintaining the advantages of a sensor with a small collection electrode. Through dedicated TCAD simulations \cite{munker2019simulations} it was found that two different process modifications could improve the charge collection and increase the radiation tolerance. These include: a design with a gap in the n-blanket implant (NGAP) and a design with an additional deep p-well implant (XDPW). These pixel flavours are produced on high-resistivity epitaxial (Epi) substrates (approximately 30 $\mu$m sensing layer on a 70-270 $\mu$m substrate) and on thick p-type Czochralski (Cz) substrates. Introducing the same sensor design on high-resistivity (3-4 kOhm) Cz substrates enables the combination of the advantages of small electrode CMOS sensors with those of a thick (100 - 300 $\mu$m) detection layer. Here the low capacitance is maintained while the signal amplitude is increased due to the thicker depleted sensor layer \cite{pernegger2023malta}. \\

\subsection{MALTA Telescope Planes}

\begin{figure}
\centering
\resizebox{0.5\textwidth}{!}{
  \includegraphics{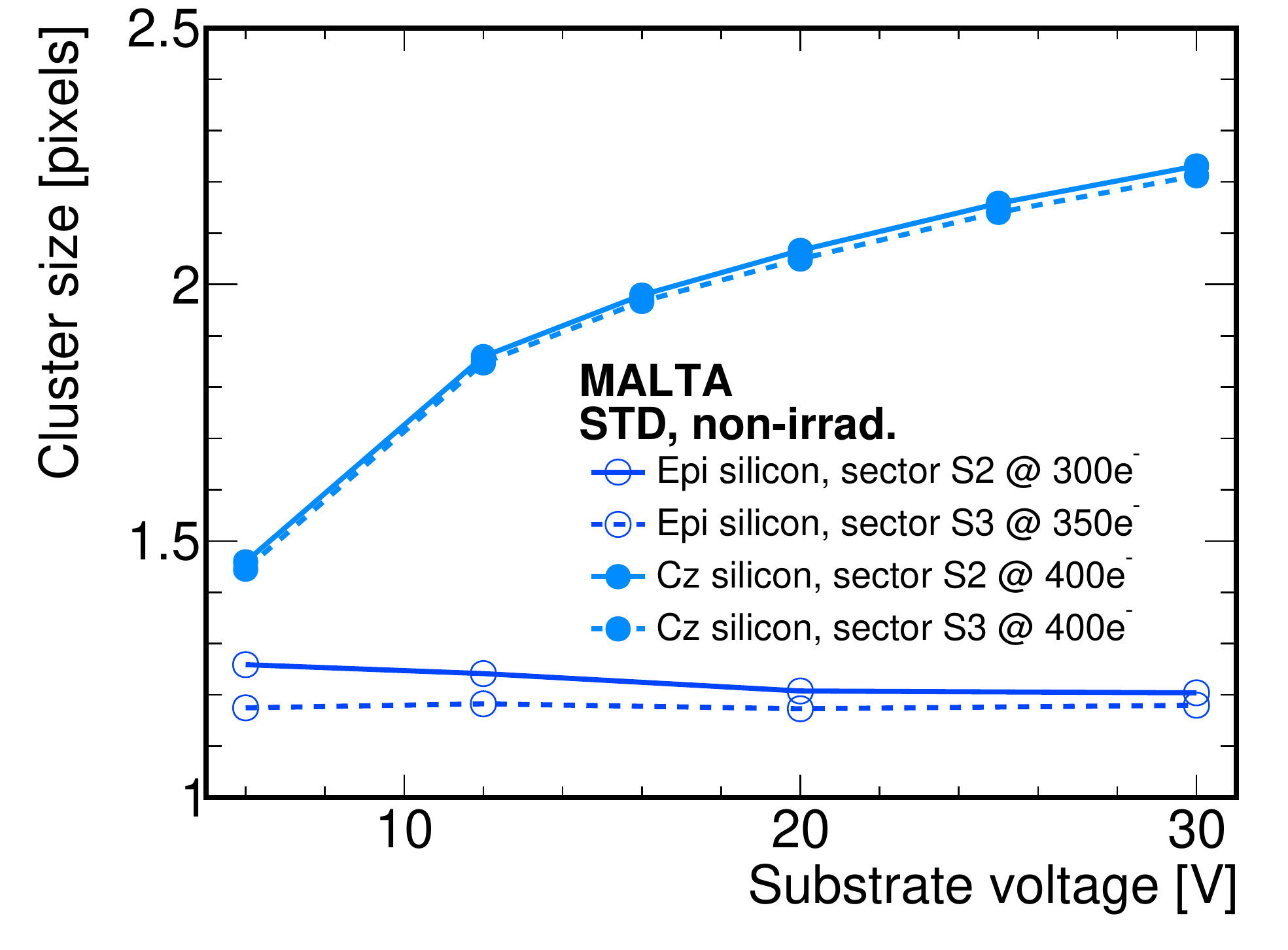}
}

\caption{Average cluster size of non-irradiated STD MALTA samples on epitaxial and Cz substrate versus reverse substrate bias. Indicated are the respective discriminator thresholds (electrons) and the corresponding sector of the pixel matrix (S2 or S3). The sectors differ in the extension of the deep p-well, with medium and maximum extension for sectors S2 and S3 respectively \cite{cardella2019malta}. }
\label{clsize}       
\end{figure}

 \noindent  It has been shown in previous test beam campaigns \cite{pernegger2023malta} that the substrate type and process modification flavour (i.e. STD, NGAP or XDPW) have great influence on the cluster size, defined in Chapter \ref{reco}. As can be observed in Figure \ref{clsize}, the cluster size for Epi sensors does not increase with substrate voltage. However, for a STD Cz sample the cluster size increases up to 2.2 pixels at 30 V reverse substrate bias due to the larger depletion depth in the thick substrate and therefore more charge is collected. This insight has been extremely valuable in designing the MALTA telescope and choosing the appropriate tracking planes, especially in order to improve its spatial resolution.\\

 \noindent Table \ref{tabel1} displays some of the key parameters of the MALTA planes that are implemented in the telescope. This includes the substrate type (Epi or Cz), the flavour (STD, NGAP, or XDPW), the thickness of the sensing layer, the bias voltage at which they are operated, and their relative position with respect to Plane 1. The position of the MALTA Cz STD planes (Plane 3 and 4) have been chosen such that they are positioned before and after the DUT(s) and are operated at their maximum bias voltage in order to guarantee the generation of large clusters.\footnote{The criterium for the maximum bias voltage is chosen such that the leakage current is minimised. } The benefits of these operating conditions on the performance of the telescope will become more evident in Chapter \ref{spatialresolution}.

\begin{table*}[!ht]
\centering
    \begin{tabular}{ |  p{4.3cm} |  p{1.3cm}  | p{1.3cm} | p{1.3cm} | p{1.3cm} | p{1.3cm} | p{1.3cm} |}
           \hline
            \multicolumn{7}{c}{MALTA Telescope Planes} \\ \hline
   \textbf{Plane}  &  \textbf{1} &  \textbf{2} &  \textbf{3} &  \textbf{4} &  \textbf{5} &  \textbf{6}    \\ \hline
 \textbf{Specifications} & & & & & & \\ \hline 

Substrate Type & Epi & Cz & Cz & Cz & Cz & Epi \\ 
Sensor Flavour &  STD & NGAP & STD & STD & NGAP & STD \\ 
Total Thickness  [$\mu$m] & 100 & 100  & 300  & 300 & 100  & 300 \\ 
Operation Voltage [V] & -6 & -6 & -30 & -30 & -6 & -6 \\

Distance [cm] & 0 & 8 & 16 & 94 & 102 & 110 \\ \hline

          \end{tabular}
 \caption{Overview of the main specifications of the MALTA telescope planes. For every telescope plane (1 - 6) the substrate type, pixel flavour, thickness, and operating voltage have been indicated. The position of the planes with respect to Plane 1 are listed. }
 \label{tabel1}
 \end{table*}

\section{Components of the Telescope}

\subsection{Mechanical Architecture}

\begin{figure}
\centering
\resizebox{0.5\textwidth}{!}{
  \includegraphics{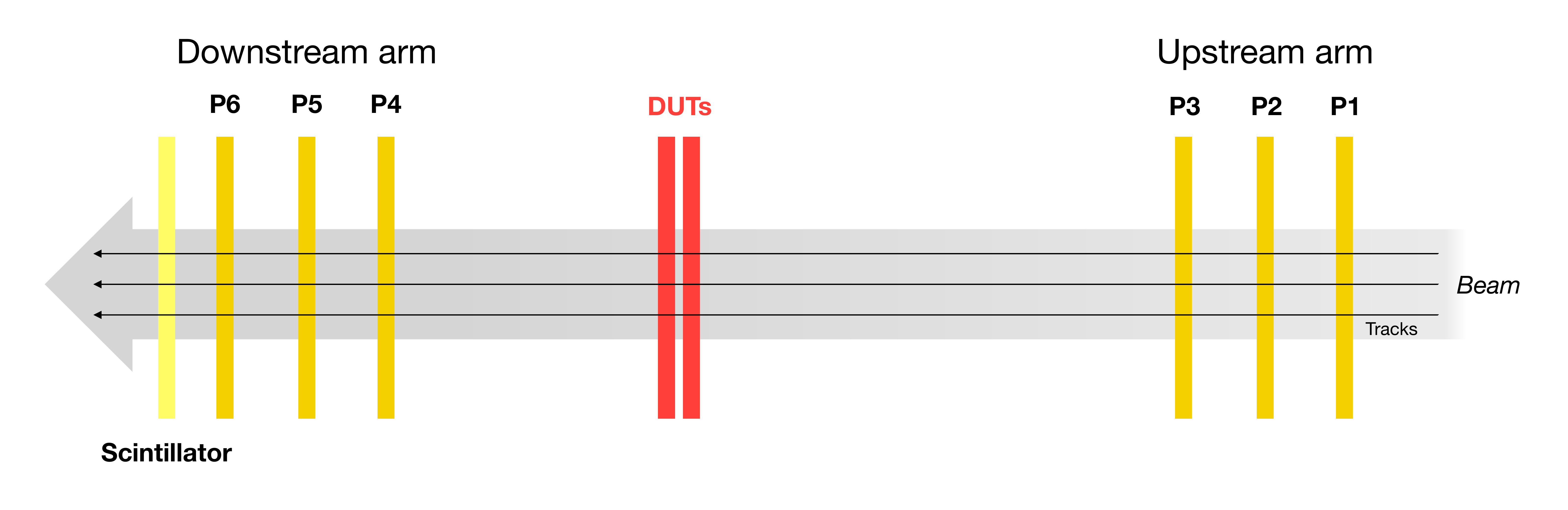}
}

\caption{ASchematic sketch of the MALTA beam telescope where the six tracking planes (P1 - P6), the device under tests (DUTs), and the scintillator are indicated.} \label{fig:telescope_diagram}
\end{figure}

A layout sketch of the MALTA beam telescope is shown in Figure~\ref{fig:telescope_diagram}. It consists of two arms placed around one or more DUTs, where each arm houses three reference planes spaced 8 cm from each other. The two arms are placed 78 cm from each other to accommodate the coldbox (explained further below). \\

\noindent The DUTs and telescope arms are held by a metal structure that is supported by two moving stages each, a STANDA micro-position and an ISEL linear stage, which can be independently positioned in both perpendicular and horizontal direction with respect to the beam. The stage that moves the arms of the telescope planes has a range of 75 cm, whereas the range of the stage holding the DUTs is 15 cm (in horizontal and perpendicular direction). Furthermore, a rotational stage allows to variate the incident angle between the beam and the DUTs. The position of the planes and the DUTs is controlled by dedicated software packages and can therefore conveniently be operated remotely. \\

\begin{figure}
\centering
\resizebox{0.5\textwidth}{!}{
  \includegraphics{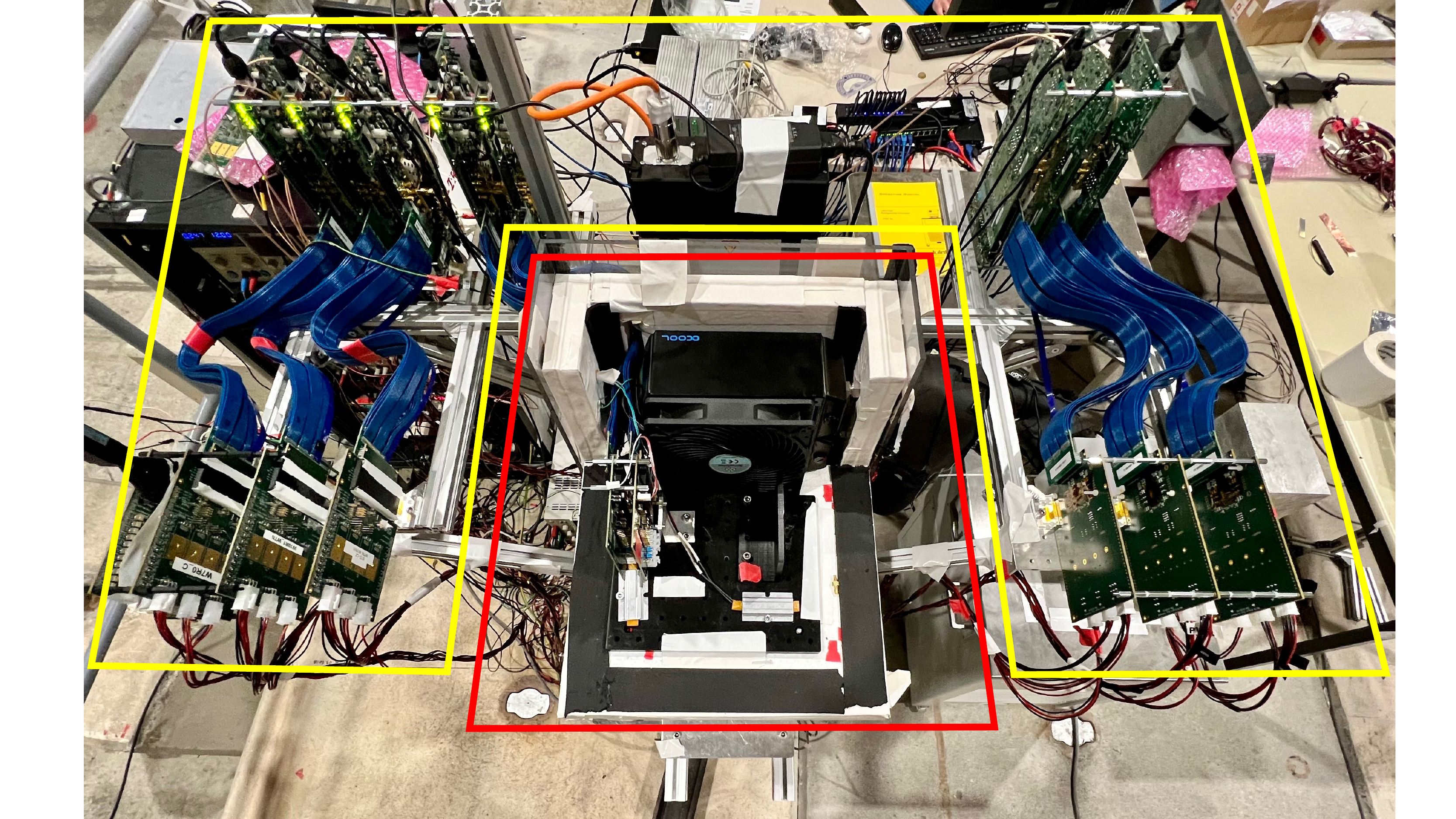}
}

\caption{Top view of the MALTA telescope. Equipment mounted on the main stage are inside yellow lines, cold box and DUTs positioned on DUT stage are inside the red lines. }
\label{fig:telescope_top_view}       
\end{figure}

\noindent The telescope relies on a JULABO FP50-HE air-cooled chiller along with a Julabo H5 thermal fluid. This cooling system is capable of reaching temperatures as low as -50$^{\circ}$C and it is employed to cool irradiated DUTs through convection. 
The DUTs can be installed inside a coldbox (46$\times$29$\times$39 cm) located on top of their linear stage, as seen in Figure \ref{fig:telescope_top_view}.
The box is completely encapsulated with a 4 cm thick expanded polystyrene layer, which insulates, in addition to the poly-methyl methacrylate walls of 5 mm, the interior. The chiller is connected to a heat exchanger and a fan to circulate the air inside the box. The usage of an external dry air source reduces the humidity in order to avoid water condensation and ice formation inside the box.
A temperature, humidity, and dew-point sensor allows for remote monitoring of the climate inside the coldbox. Through a small slit in the coldbox, the dry air pipes, power, communication, and temperature-monitoring cables are drawn in. If no cooling is required, the lid of the coldbox can be removed and the operation can proceed at ambient temperature and humidity.\\

\noindent Every tracking plane is connected to a Xilinx Kintex (KC-705) or Virtex (VC-707) commercial FPGA evaluation board using a flexible HPC FMC cable.
The FPGAs are in turn connected to the TLU through coaxial cables, further described in section \ref{trigger}. A custom framework allows the remote control of the power supplies, including TTIs (PL303QMD-P) and Keithleys (2410). This has allowed to separately control and monitor the power domains of the tracking planes and DUTs, i.e. the power supplied to the substrate (SUB), PWELL, and the analog and digital voltage domains of the chip (AVDD and DVDD). 
A network-connected Power Distribution Unit (APC AP7921B) with multiple outputs distributes the power for several parts of the telescope including the PC, linear stages, and private network switches. This allows the system to be remotely rebooted independently from the PC and the private network.

\subsection{Trigger Logic Unit}
\label{trigger}

\begin{figure}
\centering
\resizebox{0.5\textwidth}{!}{
  \includegraphics{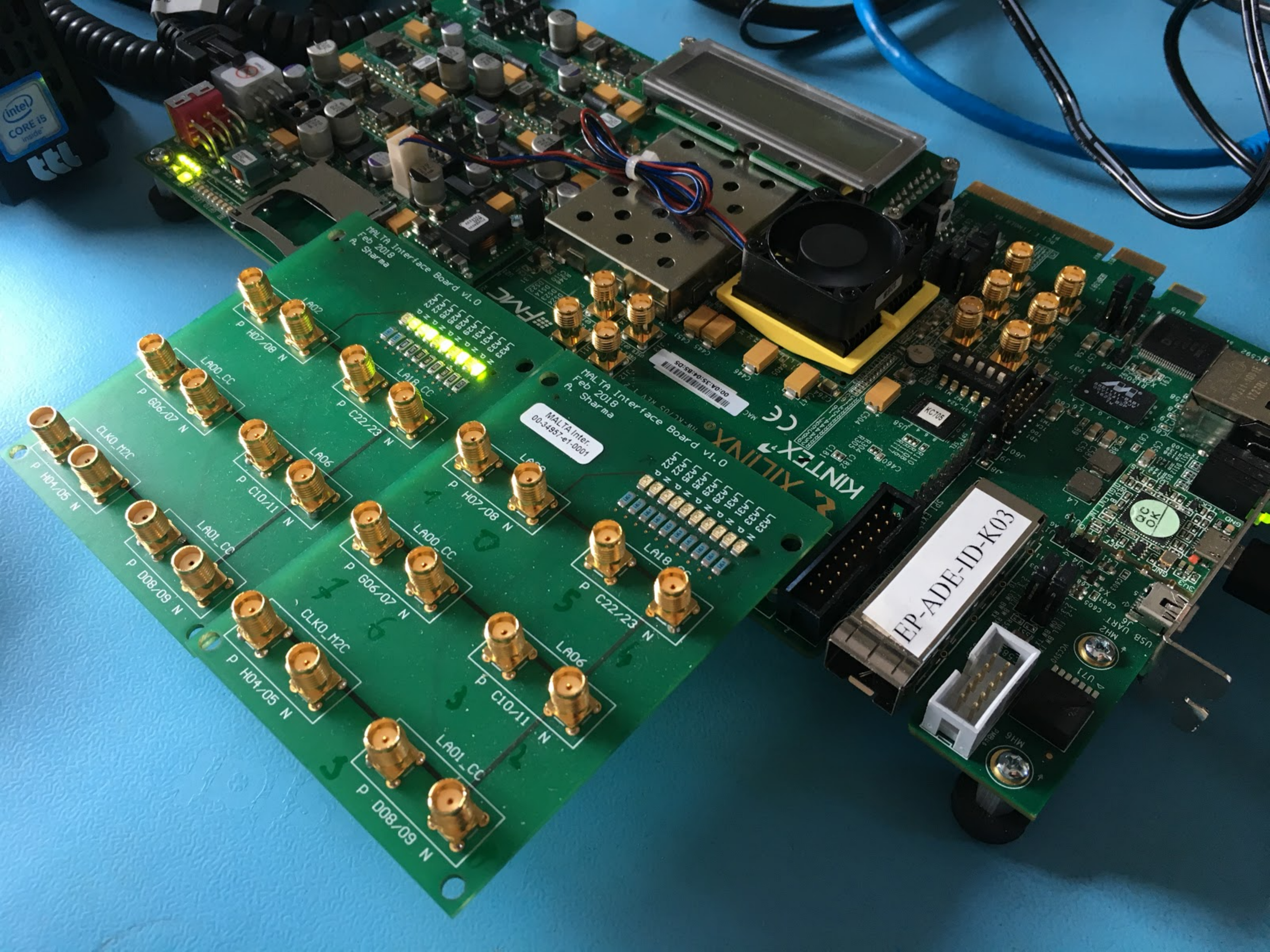}
}
\caption{SMA to FMC converter cards interface with TLU Kintex-7.}
\label{fig:TLU_setup_1}       
\end{figure}

\begin{figure}
\centering
\resizebox{0.5\textwidth}{!}{
  \includegraphics{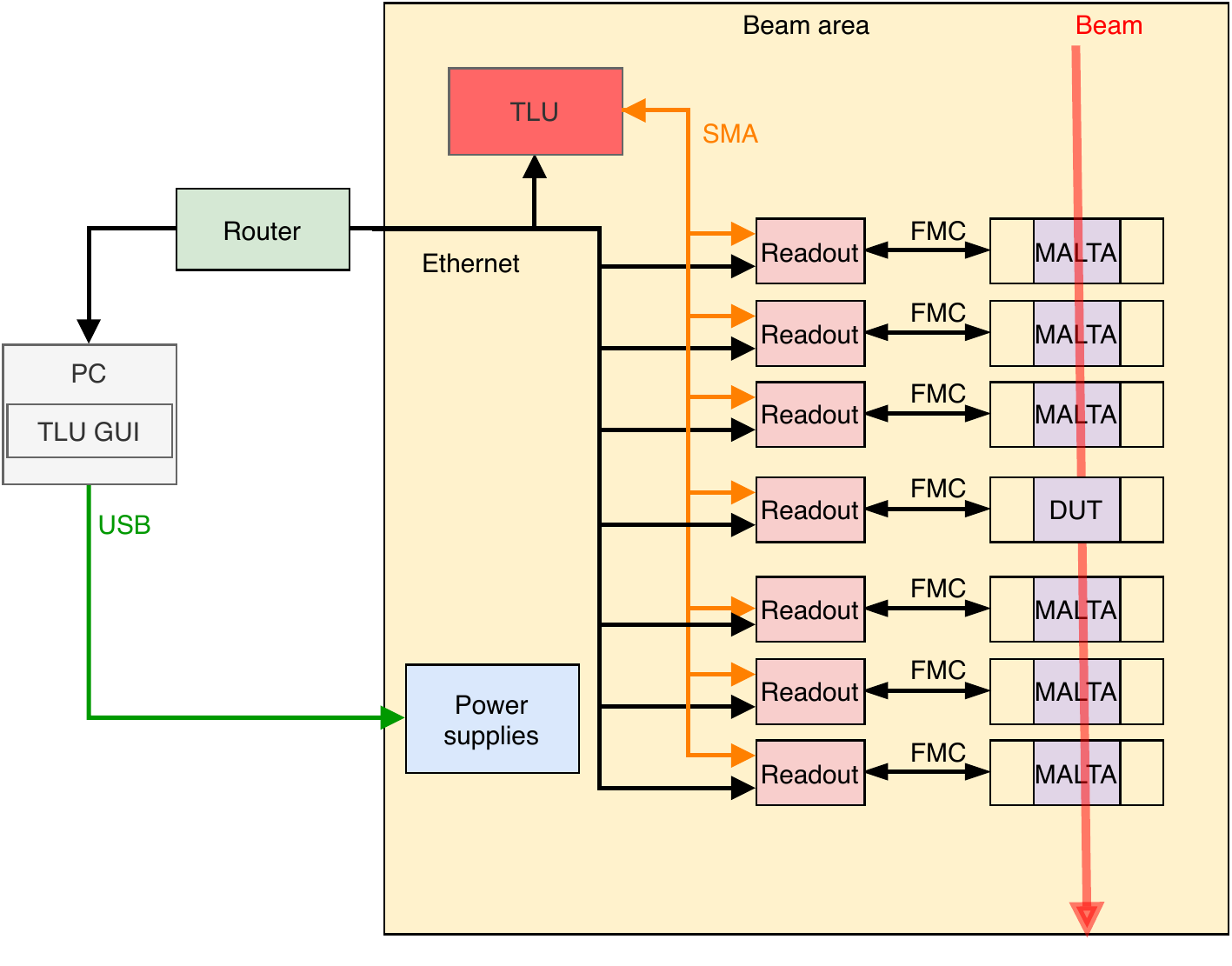}
}
\caption{Diagram of the connections between the MALTA telescope and the TLU.  }
\label{fig:TLU_setup_2}       
\end{figure}

\noindent The predecessor of the current MALTA TLU was based on Nuclear Instrumentation Modules (NIM), requiring manual configuration of the trigger logic, which made the process a complex and time-consuming task. However, FPGAs can be used to build a more versatile TLU for test beam campaigns, additionally reducing the cost and weight. The current TLU of the MALTA telescope is based on a Kintex-7 KC705 evaluation board, which is used to process the combination logic and provide online monitoring. The TLU is interfaced using SMA cables to the telescope planes and scintillator through two custom SMA-to-FMC converter cards \cite{abhipcb} for input and output signals, as seen in Figure \ref{fig:TLU_setup_1}. The Gigabit Ethernet port uses the IPbus protocol \cite{ipbus} for readout communications, control, and configuration. A micro-USB port is used for firmware programming of the FPGA through JTAG. The connections between the telescope and the TLU are shown in the Figure~\ref{fig:TLU_setup_2}.\\

\begin{figure}
\centering
\resizebox{0.5\textwidth}{!}{
  \includegraphics{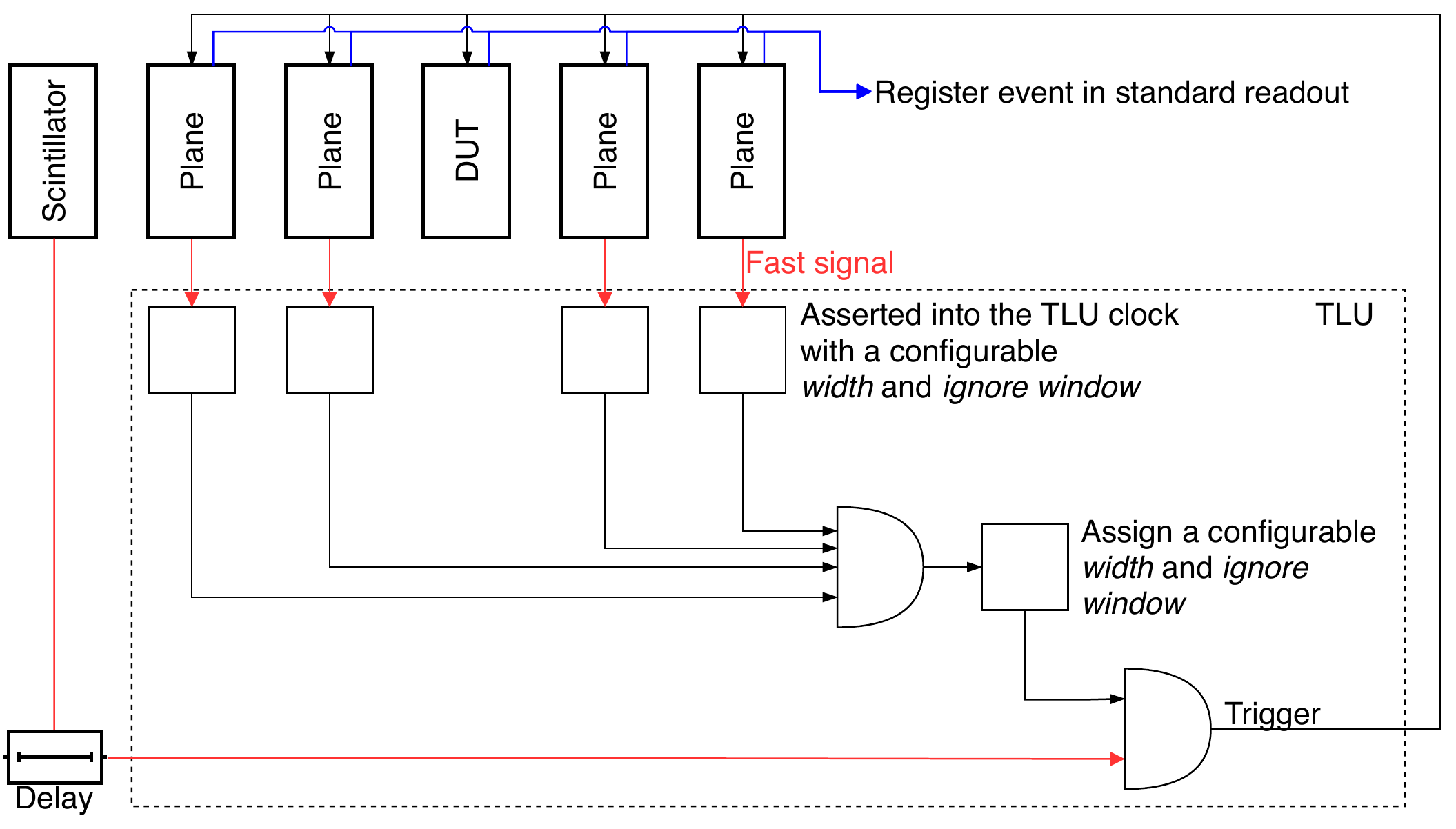}
}
\caption{Diagram of the signal processing from fast signals to trigger and data recording. The dashed black line represents the processes that occur inside the hardware of the TLU.}
\label{combination_schema}       
\end{figure}

\noindent Figure~\ref{combination_schema} shows the process that triggers the recording of events in the MALTA telescope. When a particle is detected, each MALTA plane produces two signals. The first signal is a fast signal (or hitOR), showed in red, that is sent without any processing to the TLU. The second signal is a full time stamping signal (showed in blue) which, after a defined internal delay, is sent out of the FPGA to be recorded by the standard readout. Depending on the telescope configuration, up to 4 fast MALTA signals are connected to the TLU input connectors. These signals are processed and then addressed in a combination logic to produce a combined signal. The scintillator produces a faster signal, also in red, that allows the TLU for precise timing measurements. The coincidence of MALTA planes in the combined signal has a time resolution of several ns but the fast signal of the scintillator is used as a timing reference, hence ensuring a precision in the trigger signal of a fraction of a ns. The trigger signal or Level 1 Accept (L1A) is sent from the TLU to each MALTA plane (FPGA) to trigger the recording of the full time stamping signal by the standard readout. \\

\noindent The FPGA runs a firmware divided in 3 main modules: input, coincidence logic, and output. A 320 MHz clock is generated from the FPGA internal clock for signal processing modules and logic. In the input module, as seen in Figure~\ref{combination_schema}, the asynchronous input fast signals are captured by a signal processing block into the internal clock. Each signal is transformed into a standard logic for the subsequent processing including the stretching to a programmable length and the implementation of a possible veto window to avoid too close signals. The module contains a 32-bit counter to monitor the input rate of each channel. \\

\noindent The coincidence module combines the selected individual channels in an AND gate. The width of the signal from the previous step, i.e. before the coincidence, acts as a coincidence window in the combination step. This window is necessary due to the nature of the signal from the MALTA planes on which the arrival time of the hits is proportional to the charge deposition. As such, the input signals are spread by typically 5 to 15 ns. This is observed in Figure~\ref{combination_schema}, where the L1A rate as a function of the stretched window width of the signals for three telescope planes is shown. This measurement was performed with a $^{90}$Sr source, which explains why the observed rate is lower than in test beam conditions. The larger the time window of the MALTA planes is, the greater the opportunity is for the coincidence logic to form a L1A trigger. A saturation effect can be observed after approximately 25 ns as hit signals are ignored during the long stretched processed signals.\\

\noindent The output processing module is similar to the input processing, though it allows to regulate the output signal. The capability to control the output signal length is important to interface the TLU to the devices receiving the trigger, while the veto is used to implement a maximum trigger rate. While individual chips and the TLU can theoretically perform at very high rate, the output rate of the TLU is limited to 50 kHz due to limitations of the FIFO size in the FPGA that interfaces the individual planes. 

\begin{figure}
\centering
\resizebox{0.5\textwidth}{!}{
  \includegraphics{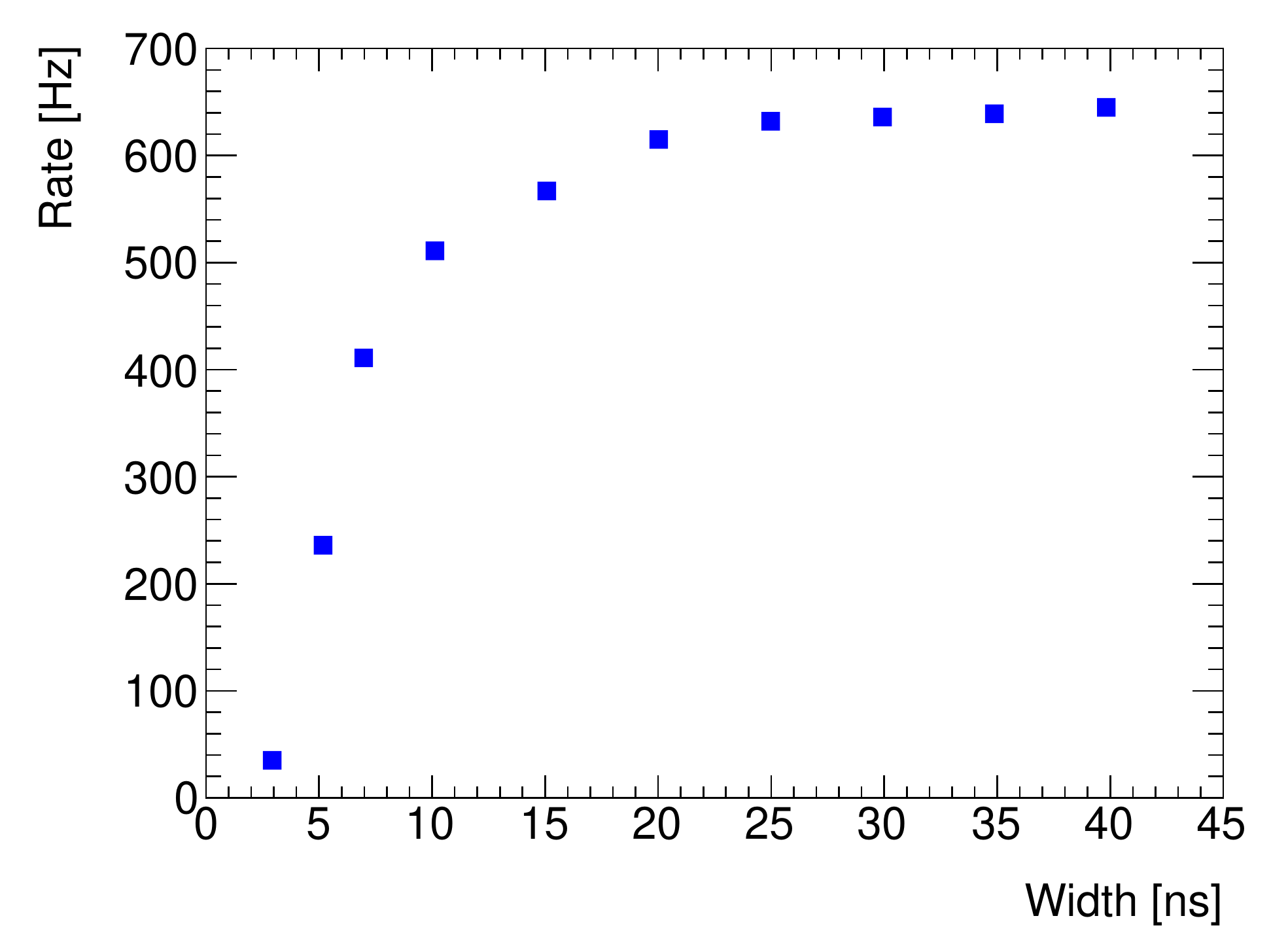}
}
\caption{Trigger rate as a function of the length assigned to the fast signals to the combination logic. Measurement was done with low energy electrons from $^{90}$Sr $\beta$-decay. }
\label{TLU_rateVSwidth}       
\end{figure}

\subsection{Data Aquisition}

\begin{figure}
\centering
\resizebox{0.5\textwidth}{!}{
  \includegraphics{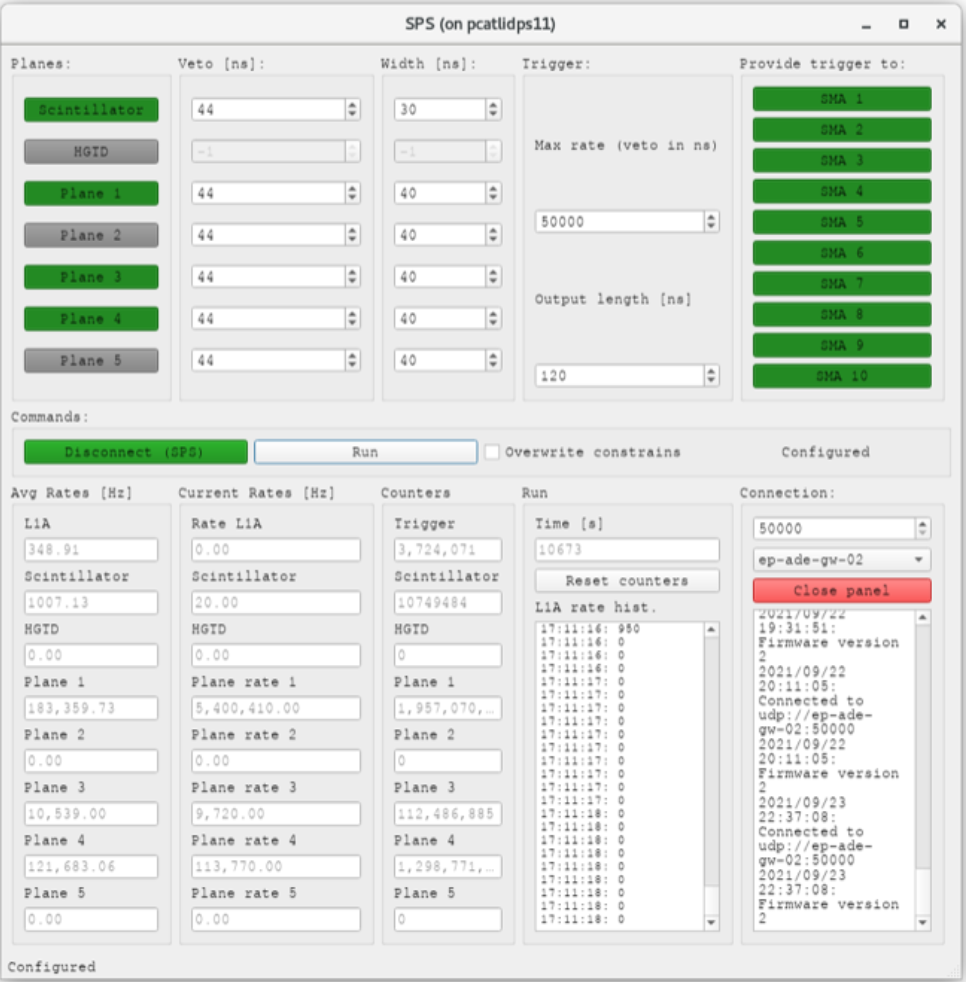}
}
\caption{Screenshot of the in-time trigger logic unit GUI during data-taking. The GUI allows the user to start and stop a run, select on which plane (or scintillator) to trigger on, the maximum rate that is allowed, the output length of the trigger signal, and displays in-time rate for every plane.}
\label{fig:TLU_UI}       
\end{figure}

\noindent The DAQ process of the TLU is implemented inside the FPGA as a Finite State Machine (FSM) that is controlled remotely via IPbus. A Command Line Interface (CLI) and Graphical User Interface (GUI) are available for this, the latter shown in Figure~\ref{fig:TLU_UI}. The baseline is a C++ class responsible of communication with the FPGA. The upper panels allow the selection of trigger planes, maximum allowed rate (veto), the output signal width, L1A and connection settings. The middle panel allows the user to start and stop a data-taking run. The lower panel is used to monitor the number of triggers of the respective planes and L1A.  \\

\begin{figure*}
\centering
\resizebox{1.0\textwidth}{!}{
  \includegraphics{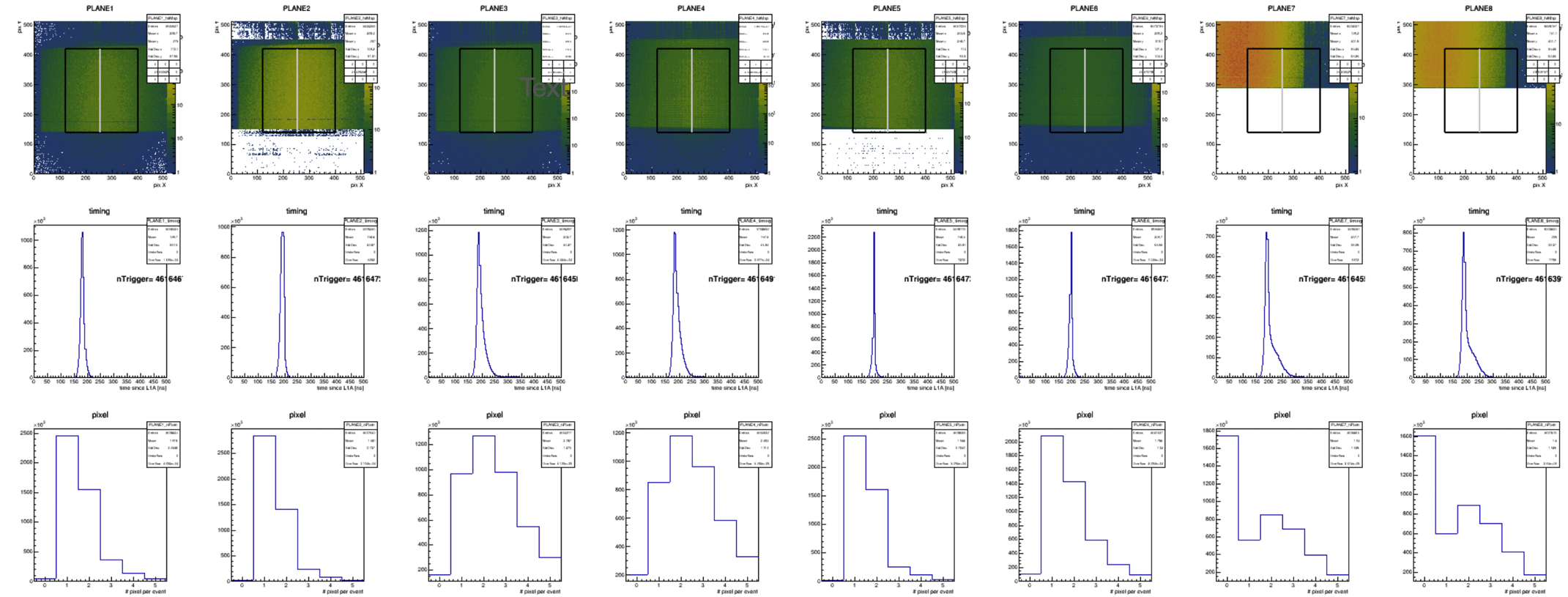}
}
\caption{Screenshot of the in-time telescope data acquisition window during data-taking. From left to right the six tracking planes are indicated, with the last two columns representing the two DUTs. The top plot shows the hit-map, the middle plot shows the time of arrival with respect to the trigger, and the bottom shows the hit multiplicity. }
\label{fig:DAQ_screenshot}       
\end{figure*}

\noindent Figure~\ref{fig:DAQ_screenshot} shows the in-time data acquisition window during data-taking. Every column represents a tracking plane (Plane 1 - 6), while the last two planes represent the two DUTs (Plane 7 and 8). The top row shows the hit map of each plane, which allows additionally to visualise a ROI. The middle row shows the number of hits as a function of time of arrival since the L1A signal with respect to the trigger scintillator. The scale of the plot (500 ns) represents the readout window after the L1A. The bottom row shows the hit multiplicity histogram, i.e. the number of hits as a function of the number of pixel per event. The realtime monitoring plots allow for quick feedback on and fast assessment of the DUT performance.

\section{Reconstruction and Offline Analysis}
\label{reco}

\noindent Precise knowledge of the positions of the detector planes is required to guarantee a high track quality and correct association with DUT clusters. The MALTA telescope is mechanically built with as much precision as possible. However, as robust mechanical frameworks have a limited spatial accuracy when setting the positions of the planes, offline measurements of their deviation can be accounted for during the software alignment. For the alignment, track reconstruction, and the offline analysis of test beam data the software package Proteus is used \cite{kiehn}. Proteus software takes raw data in the form of hits per event and groups them into clusters, where a cluster is defined as the geometrical average of the adjoining hit positions. It can also provide track reconstruction by finding tracks from clusters on the tracking planes and it calculates the optimal track parameters on selected planes. Finally, Proteus provides the user with output data for further offline analysis. The sequence of alignment, reconstruction, and analysis with Proteus are discussed in more detail below. Using these modular steps, the performance of the telescope planes is discussed in the final section of this chapter. 

\subsection{Alignment}

\noindent The alignment step accounts for possible misalignment from the nominal telescope description. This nominal telescope position uses Plane 1 (upstream of the beam) as the origin of a global coordinate system to which the other planes downstream are referenced towards. In total, three rotational degrees (around x-, y- and z-axis) and two translational degrees of freedom (in x- and y-direction) per tracking plane are considered. During the first step of the alignment, i.e. the coarse alignment, the hit correlation distribution between consecutive planes is calculated. Assuming a parallel beam, the detector misalignment perpendicular to the beam is inferred by the correlation offset, while deviations from the calculated slope are directly proportional to deviations in the rotational direction. The next alignment step, i.e. fine alignment algorithm, uses the unbiased residuals to iteratively perform a $\chi^2$-minimization. This fine alignment is motivated by the fact that the track position is influenced by the position of the given plane being aligned. The fine alignment runs until the  $\chi^2$ is minimal and convergence is achieved, where any calculated shift is used in the initial geometry. 

\subsection{Track Fitting}

\begin{figure}
\centering
\resizebox{0.5\textwidth}{!}{
  \includegraphics{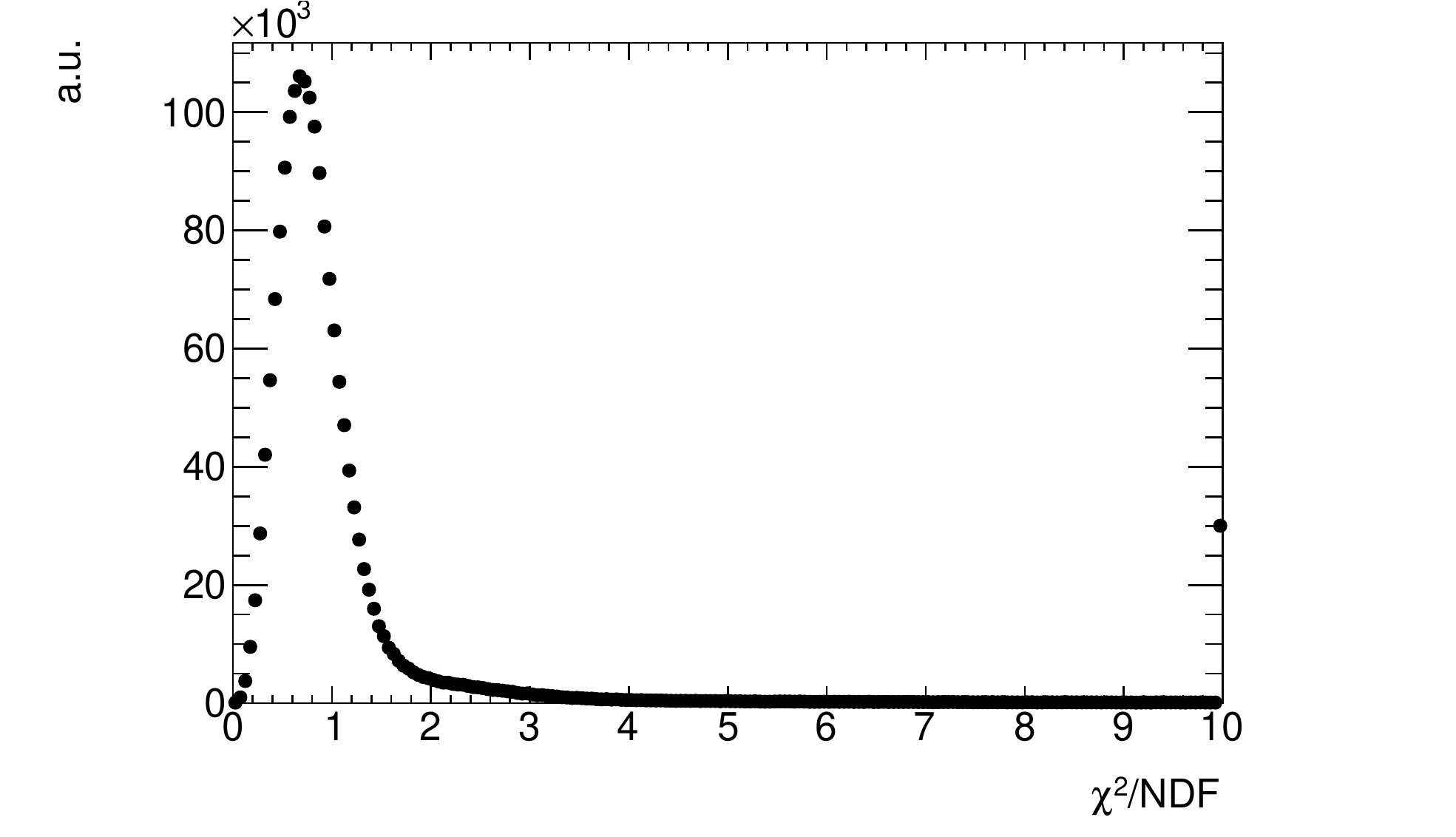}
}
\caption{Distribution of the $\chi^2$ divided by NDF after the alignment and track fitting steps of Proteus. The last bin includes an overflow.}
\label{Chi2}       
\end{figure}

\noindent The alignment step is completed when the positions of the telescope planes are well-defined, after which the tracks can be reconstructed. By using a seed cluster, the tracking algorithm searches for clusters on the consecutive telescope planes. Depending on the beam type, an angle in which this search is performed can be chosen in order to compensate for potential scattering. In case multiple clusters are found within the region the algorithm searches, the track search will split and continue in both directions. Only tracks that contain the largest number of associated cluster are kept, after which a $\chi^2$-cut is applied to filter out tracks of bad quality, for instance due to multiple scattering or nuclear interaction with the telescope planes. In Figure \ref{Chi2} the distribution of the $\chi^2$ divided by the number of degrees of freedom (NDF) is shown. The distribution shows a peak at one after the alignment and track fitting routines, indicating good fitting results. The $\chi^2$/NDF<10 requirement is applied to tracks. It allows for some buffer during automated analysis for unexpected misalignment, without significantly affecting the performance in terms of efficiency nor timing resolution. 

\subsection{DUT Cluster Matching}

\noindent Finally, hit clusters on the DUT need to be associated to a track. In order to do this, the fitted track is extrapolated to the DUT and the matched clusters are restricted to a user-defined distance. The minimum value of this distance is dependent on the spatial resolution of the DUT. In our studies this distance is set to 2.5 times the size of the pixel pitch (36.4 $\mu$m), i.e. 80 $\mu$m. When the matched clusters on the DUT are found, it is possible to define the DUT efficiency as the number of tracks with a matched cluster on the DUT over the total number of reconstructed tracks. 

\subsection{Characterisation of Telescope Planes}

\noindent The telescope planes were characterised during test beam measurements through dedicated runs where one plane at a time was selected as the DUT. Due to the flexibility of the TLU, other planes except the DUT could be selected as triggering planes. As a result the average efficiency and cluster size of these planes could be measured. The results of these measurements are presented in Table \ref{tabel2}. Additionally, in this tabel the average number of hits per event (within a 500 ns acquisition window) is shown, a variable that can be monitored during real-time data-taking (Figure \ref{fig:DAQ_screenshot}).

\begin{table*}[!ht]
\centering
    \begin{tabular}{ |  p{5cm} |  p{1.6cm}  | p{1.6cm} | p{1.6cm} | p{1.6cm} | p{1.6cm} | p{1.6cm} |}
           \hline
            \multicolumn{7}{c}{Characterisation of Telescope Planes} \\ \hline
   \textbf{Plane}  &  \textbf{1} &  \textbf{2} &  \textbf{3} &  \textbf{4} &  \textbf{5} &  \textbf{6}    \\ \hline
&  &  &  &  &  &  \\ \hline 
Average Efficiency [\%] & 97.4 $\pm$0.1 & 96.9 $\pm$0.1 & 99.2 $\pm$0.1& 99.0 $\pm$0.1 & 95.0 $\pm$0.1 & 94.0 $\pm$0.1  \\  \hline 
Average number of hits per event & 1.7 $\pm$0.9 & 1.8 $\pm$1.0  & 2.2 $\pm$1.1 & 2.5 $\pm$1.2 &  1.7 $\pm$0.9  &  1.6 $\pm$0.9 \\ \hline 

Average Cluster Size [pixels] & 1.1 $\pm$0.3 & 1.1 $\pm$0.4 & 1.7 $\pm$0.8 & 1.9 $\pm$0.9 & 1.1 $\pm$0.3 & 1.1 $\pm$0.4 \\  \hline

          \end{tabular}
 \caption{Overview of the characterisation of the MALTA telescope planes. For every telescope plane (plane 1 - 6) the average efficiency, average number of hits per event within a 500 ns acquisition window, and average cluster size are listed. The error for the average efficiency is expressed as the statistical uncertainty. The error on the latter two parameters (average number of hits per event and cluster size) is expressed as the Root Mean Square (RMS) of the relative distributions.}
 \label{tabel2}
 \end{table*}

\section{Spatial Resolution}
\label{spatialresolution}

To evaluate the spatial telescope resolution first an analytical estimate~\cite{telescopeOptimizer} that takes into account the planes' position, the radiation length and intrinsic resolution of the sensors and the beam energy are used.
For sufficiently high momentum of the incident particles the latter parameter does not have an impact on the resolution. This is the case of the following estimation where the charged hadron beam used at SPS has an energy greater than $\mathcal{O}(100)$~GeV.
The intrinsic resolution of the sensor is set to $\sigma_{int} = 10.5~\mu$m assuming a uniform distribution on the pixel pitch. The radiation length of the sensor is about 0.1\% for the planes with samples $100~\mu$m thick and 0.3\% for those of $300~\mu$m.\\

\noindent Tracks are reconstructed with a linear fit of the clusters position based on the $\chi^2$ minimisation method. This provides the measurement of the several tracks parameters and the projection of the intercepts on the DUT. The uncertainty of the latter gives an estimation of the spatial resolution. Only events where exactly one track is reconstructed fulfilling the aforementioned fit quality criteria are selected. The RMS of the estimated resolutions is assigned as the uncertainty of the central value.
 The General Broken Lines (GBL) algorithm~\cite{Kleinwort:2012np} was also tested to account for Coulomb scattering due to the material interaction of the beam with the telescope planes and the DUTs. Results are found to be compatible with the linear fit regression. \\ 
 
\begin{figure*}
\centering
\resizebox{0.95\textwidth}{!}{
  \includegraphics{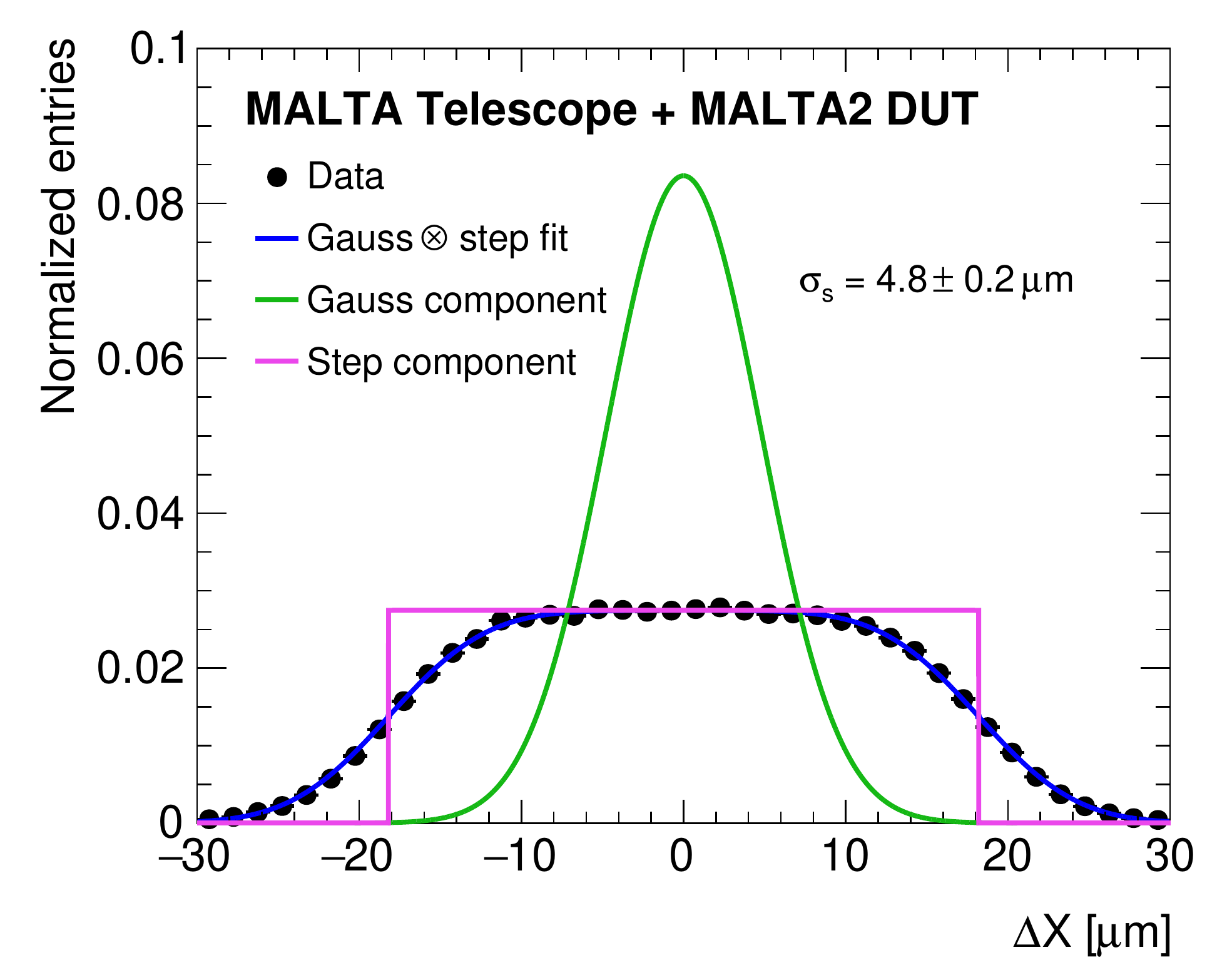}
  \includegraphics{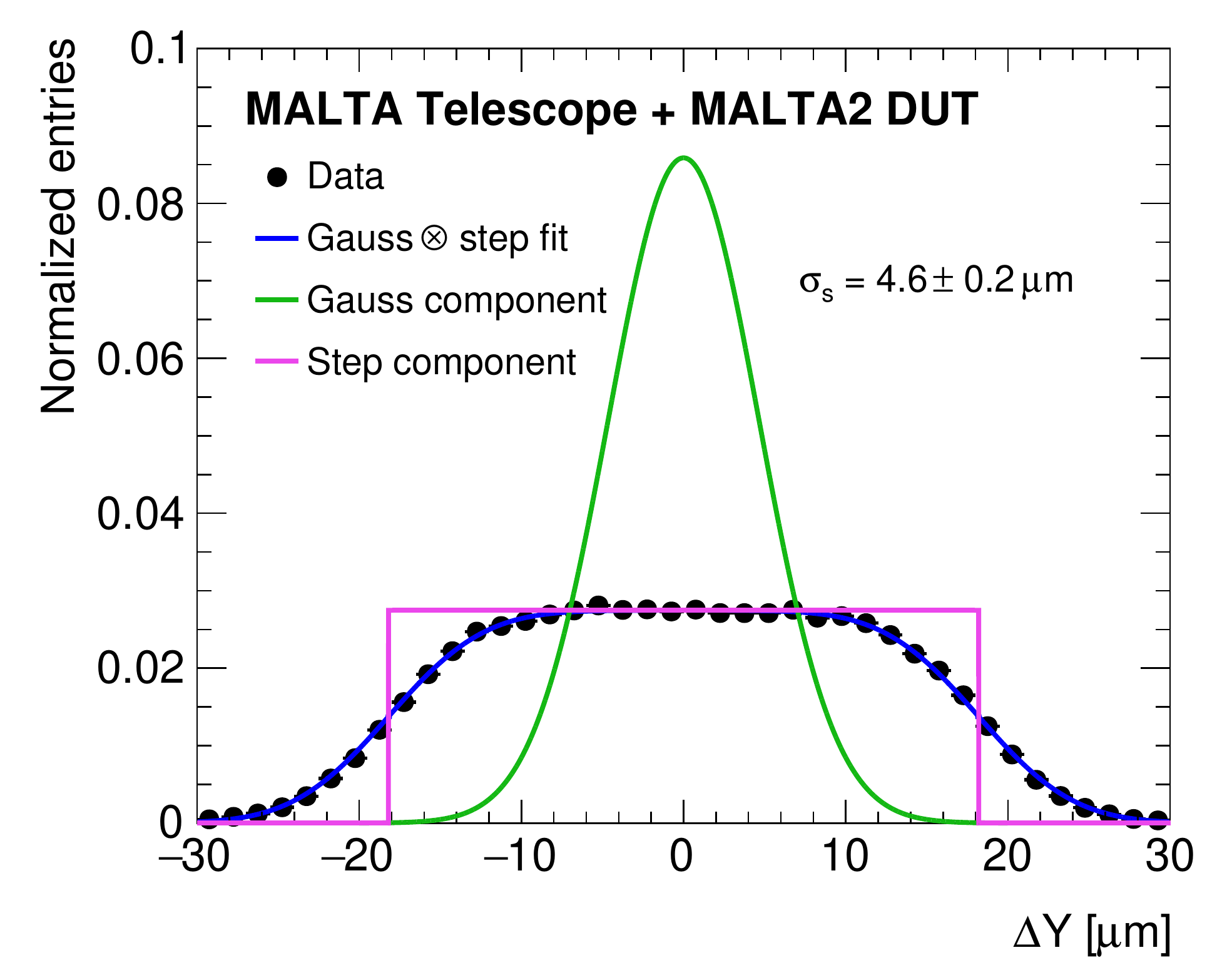}
}
\caption{Distribution of the residuals on the $X$ (left) and $Y$ (right) directions between the fastest sensor hit and the track intercept. The fit result of a convolution of a Gaussian with a two-sided step distribution is shown (blue) together with the Gaussian (green) and step function (magenta) components.\label{fig:residuals}}    
\end{figure*}

 \noindent An alternative approach adopted to evaluate the tracking resolution is based on the distance from the track intercept and the fastest hit of the closest reconstructed cluster of a DUT. 
At first, a residual mis-alignment correction is applied to data by subtracting from each calculated distance in the pixels its average obtained in the corresponding position in the chip matrix.
The track spatial resolution of the telescope can be described by a Gaussian function whereas the probability of hitting the sensor is assumed uniform along the pixel pitch. Hence, the convolution of a Gaussian and a step function is used to perform a fit on the residual on the two transversal coordinates $X$ and $Y$ of the DUT, as shown in Figure~\ref{fig:residuals}. The resulting spatial resolution is extracted from the width of the Gaussian distribution.\\

\begin{figure}
\centering
\resizebox{0.45\textwidth}{!}{
  \includegraphics{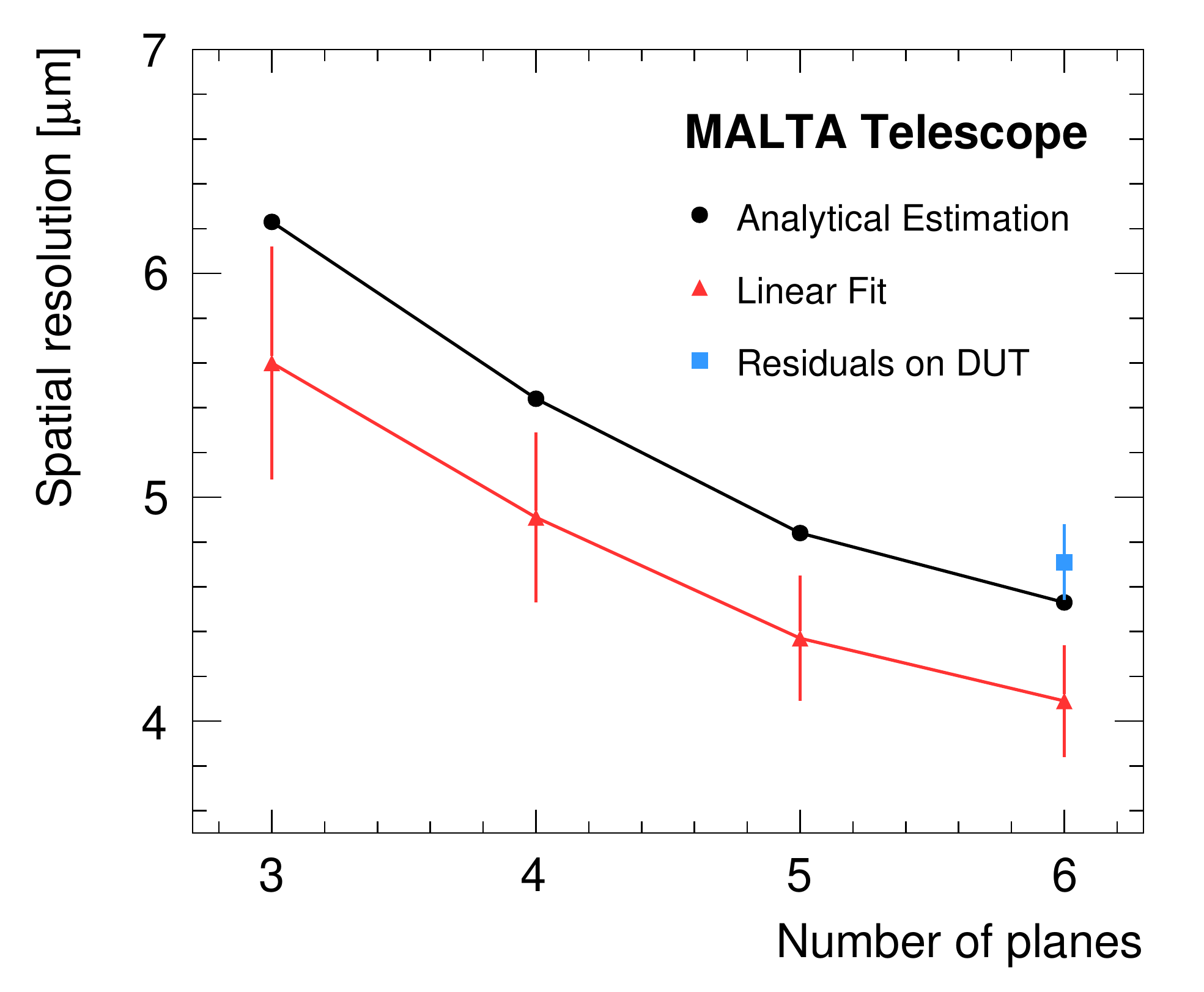}
}
\caption{Telescope resolution as a function of the number of planes considered. The analytical estimation (black) is compared with the measurements based on the linear fit (red). The total resolution is also extracted from the fit of the convolution of a Gaussian with a two-sided step function on the distribution of the residuals on a DUT (blue).\label{fig:telescope_resolution_summary}}    
\end{figure}

\noindent The analytical description of the spatial resolution on the DUT position as a function of the number of telescope planes is shown in Figure~\ref{fig:telescope_resolution_summary} together with the estimation given by the track parameters obtained from the linear fit. The set of planes chosen in the several configurations provides the best measured resolution for the fixed subset of samples. These always include the planes closer to the DUT, having a cluster size greater than one, hence improving the spatial resolution. The comparison highlights how the cluster size, not taken into account in the simulation, affects the spatial resolution improving the predictions by about 10\% for all the tested configurations.
The measured spatial resolution based on the linear regression approach of the full telescope is $\sigma_{s} = 4.1\pm 0.2  ~\mu$m. 
The average spatial resolution extracted from the Gaussian fit of the two transversal components
 is also shown and corresponds to $\sigma_{s} = 4.7\pm 0.2  ~\mu$m. This latter approach provides a larger estimate if compared with the analytical description and the linear regression measurements. This is attributed to the effects of inhomogeneity and time resolution at the edges of the DUT that are not taken into account in the fit and inflate the width of the Gaussian function.

\section{Timing Resolution}
\label{timing}

The timing performance of the MALTA tracking planes has been measured extensively during the past test beam campaigns. In order to describe the timing response of the telescope planes, several intrinsic and extrinsic effects need to be accounted for. One of these systematic intrinsic effects is the signal propagation, i.e. the time required to reach the periphery along the column direction in the pixel matrix. This effect consists of a contribution of the pulse propagation to the pixel and a contribution from the hit propagation to the periphery. As this effect has a linear behaviour, a dedicated propagation correction is applied and integrated over the whole pixel matrix. An extrinsic effect that is taken into account is the jitter originating from the scintillator, approximately 0.5 ns, and from the FPGA. The FPGA latches the trigger with the 320 MHz clock, which adds a 3.125 ns per-event jitter. Therefore, the contribution of the FPGA to the timing distribution is approximated at 3.125/$\sqrt{12}$ $\sim$ 0.9 ns. It should be pointed out that due to the fact that the trigger is internally latched within the FPGA, a lot of flexibility and adaptability to any trigger source is gained within the set-up. \\

\begin{figure}
\centering
\resizebox{0.39\textwidth}{!}{
  \includegraphics{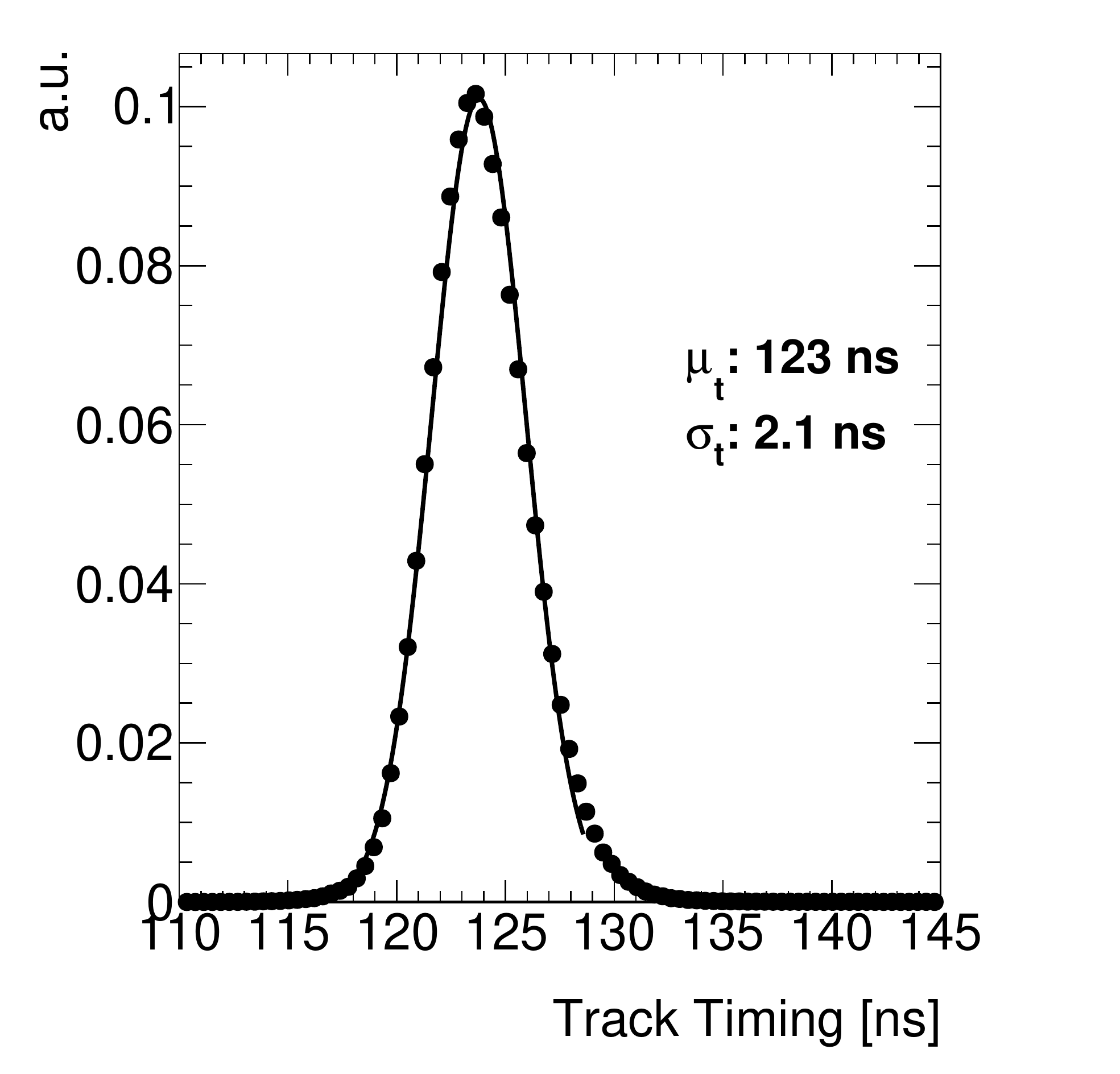}
}
\caption{Averaged timing distribution of the six MALTA tracking planes with respect to the scintillator reference. The quoted standard deviation and mean result from fitting a Gaussian to the core of the distribution. Measurements were done with a 180 GeV hadron beam at SPS at CERN in 2022. }
\label{tracktiming}
\end{figure}

\noindent Figure \ref{tracktiming} shows the track timing, where the time of arrival of the fastest hit in the cluster with respect to the scintillator for the six tracking planes is averaged to a single timing distribution. The RMS of the timing distribution of all six telescope planes were found to be compatible. The standard deviation extracted from a Gaussian fit performed on the track timing distribution provides an estimation of the timing resolution of $\sigma_{t}=2.1$ ns. The excellent timing performance allows to separate the individual contributions of the track, even in high rate beam settings (> 4$\times$10$^6$ particles per spill).

\section{Device Under Test Integration}
\label{dutintegration}

Ultimately, a telescope is dedicated to the characterisation of R\&D prototypes. As upcoming colliders and physics experiments will become more demanding, the next generation detectors will need to be able to operate at high rate and fluence conditions. The flexibility of the MALTA telescope has allowed to test a wide variety of detector prototypes during the test beam campaigns of 2021 and 2022, which provided relevant insights and results for future LHC experiments and upgrade programs. This section will highlight results that were achieved with two different types of DUTs and their respective integration in the MALTA telescope.

\subsection{MALTA2: Latest Radiation-Hard Full Scale MALTA Sensor}

The next prototype of the MALTA family, MALTA2, had a dedicated test beam campaign in the MALTA telescope starting in May 2021 until November 2021 and did an equal lengthy campaign in 2022. During this test beam campaign, the radiation hardness and timing performance of the second generation sensors was the focus point. Furthermore, post-processed MALTA2 samples were extensively tested. \\

\noindent Analogous to its small-scale demonstrator, the Mini-MALTA \cite{dyndal2020mini}, MALTA2 has two front-end designs implemented: the standard MALTA design and a cascode design. The size of selected transistors was increased to reduce the Random Telegraph Signal (RTS) noise, which opens up the possibility to operate at lower threshold and therefore reach higher efficiencies \cite{piro20221,van2022radiation}. The matrix of MALTA2 consists of 224$\times$512 pixels, with equal pixel size of the original MALTA (36.4 $\mu$m). Effectively, the active area of the chip corresponds to $\sim$18.33 mm$^2$. Similar to MALTA, this second prototype is fabricated on both Epi and Cz substrate and both the NGAP and XDPW pixel flavors are implemented. \\

\begin{figure}
\centering
\resizebox{0.45\textwidth}{!}{
  \includegraphics{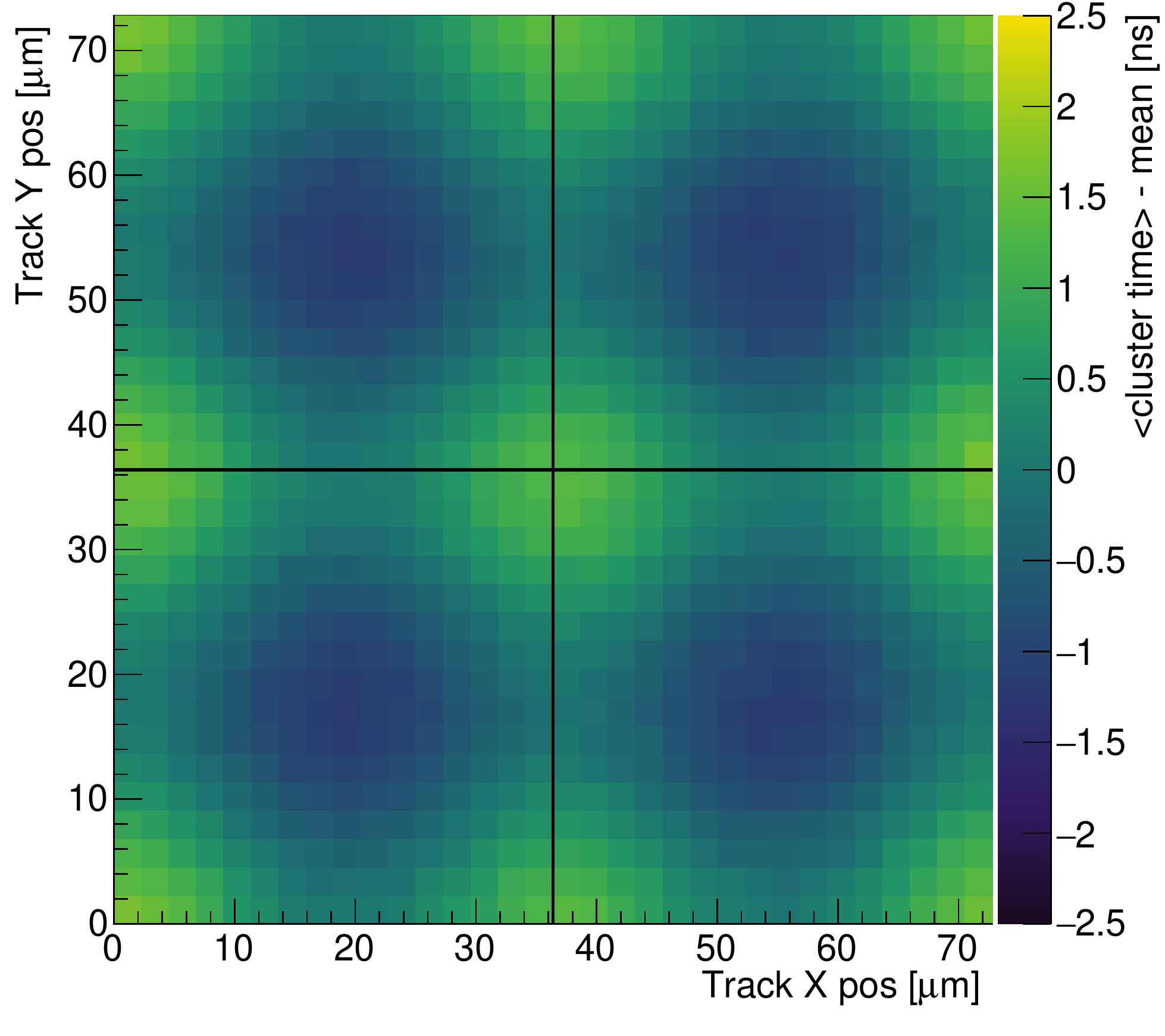}
}
\caption{In-pixel timing projected over a 2x2 pixel matrix for a MALTA2 Epi, XDPW, 100 $\mu$m thick, low doping of n-blanket, at -6 V SUB bias and -6 V PWELL bias. Colour scale indicates the difference in timing of the leading hit in the cluster and the average timing over the entire matrix. Threshold corresponds to 130 electrons. Measurements were done with a 180 GeV hadron beam at SPS at CERN in 2021.}
\label{time}
\end{figure}

\noindent The benefits of testing the MALTA2 sensor in the MALTA telescope become evident from Figure \ref{time}. This figure shows a projection of the difference between the time of arrival of the leading hit in a pixel cluster and the average arrival time of signals over the entire chip projected on a 2$\times$2 pixel matrix. As one can observer from the plot, a difference of 2-3 ns is observed between signals originated from the pixel center and the signals from the corners  \cite{gustavino2022timing}. Here, the excellent resolution of the telescope allows for fine observation of these effects between pixels and charge sharing are useful insights for future simulation work. These results have prevailed that the MALTA telescope has allowed for precise hit timing for DUTs of the order of $\mathcal{O}({ns})$.

\subsection{Carbon-Enriched Low Gain Avalanche Detectors for the ATLAS High Granularity Timing Detector}

Low Gain Avalanche Detectors (LGAD) have been extensively studied within the context of the High Granularity Timing Detector in the ATLAS experiment, which will be added during the Phase-II upgrade of the High-Luminosity Phase of the LHC \cite{upgradeHGTD}. The LGAD sensors are required to achieve at a maximum fluence of 2.5$\times$10$^{15}$ 1 MeV n$_{eq}/{cm}^2$, among other requirements, a timing resolution of 50 ps per hit, a hit efficiency of 97\% and a collected charge $> 4 $~fC at the start of their operation. During the summer of 2021, a dedicated test beam campaign with the MALTA telescope was performed using irradiated (up to 2.5$\times$10$^{15}$ 1 MeV n$_{eq}/{cm}^2$) LGAD sensors with a carbon enriched gain layer in order to improve the radiation hardness \cite{lgads}. The MALTA telescope was used to track the incident charged particles and provide the position of the incoming particle in the frame of each DUT. Below, the integration of LGAD sensors into the MALTA telescope is described.  \\

\noindent The LGAD sensors, manufactured by Fondazione Bruno Kessler (FBK) and Institute of High Energy Physics (IHEP), are mounted on a custom readout board with integrated amplification stages to enhance their signal. A four-channels oscilloscope is used to sample the waveform from the DUTs and initiates its DAQ by the trigger system. The firmware of the TLU is modified such that it only uses the accept state from the oscilloscope, where only events triggered and accepted are recorded.  In order to trigger on a particle, the second MALTA tracking plane is used in coincidence with the scintillator placed behind tracking Plane 6. If the two sensors record a signal, the TLU records the telescope data from all six planes and the waveforms from both the DUT and a second LGAD that is used as a timing reference (timing resolution of $\sim$55 ps). Finally, the track reconstruction is performed with the software package Proteus, as described in section \ref{reco}. The tracks that are reconstructed are extrapolated onto the plane of the DUT, taking into account multiple scattering by using the GBL algorithm (discussed in section \ref{spatialresolution}). \\

\begin{figure}
\centering
\resizebox{0.5\textwidth}{!}{
  \includegraphics{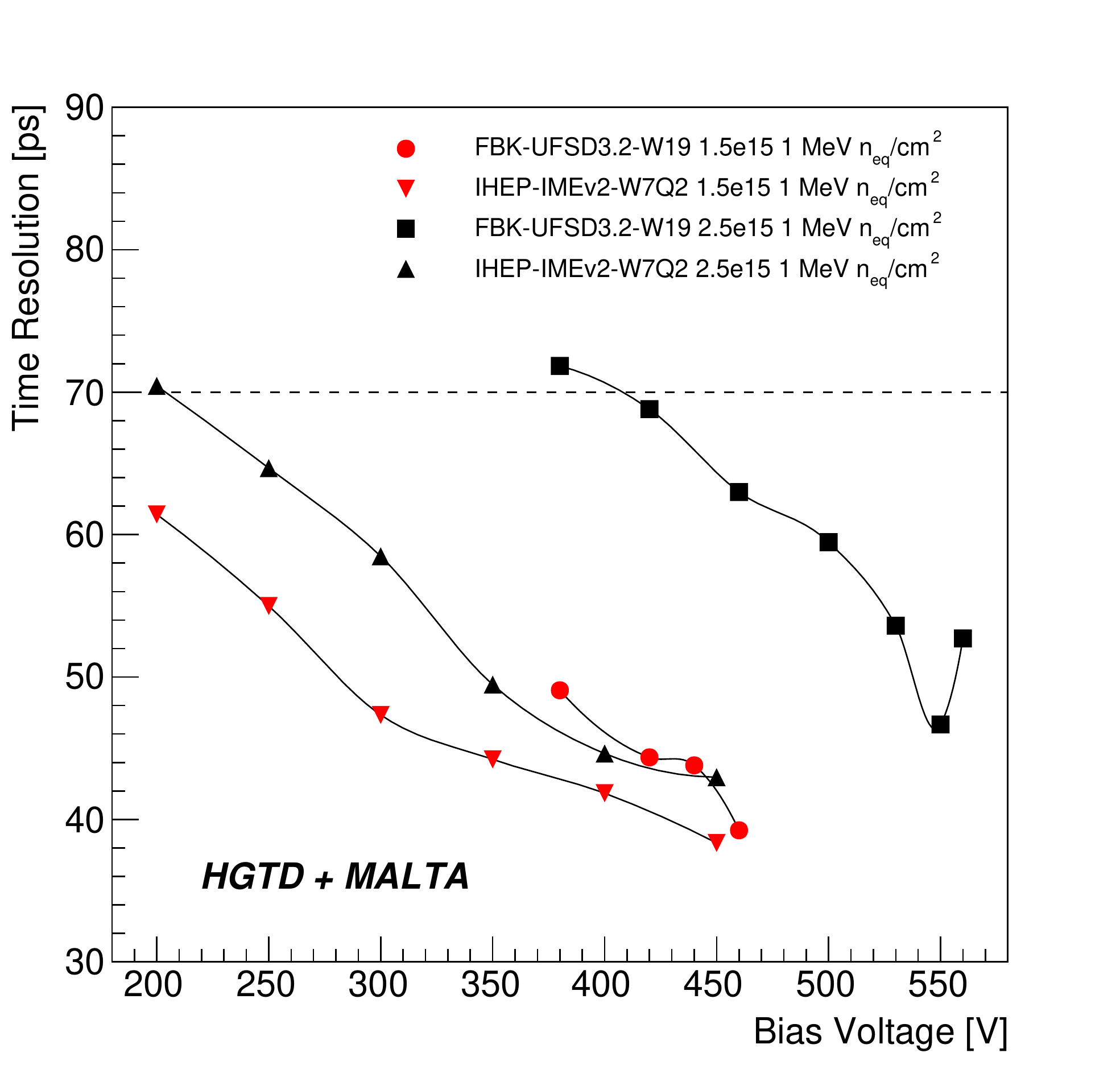}
}
\caption{Time resolution as a function of bas voltage for the HGTD test beam campaign in the MALTA telescope. The dashed lines corresponds to the minimum requirements for the future HGTD. The results are shown for different single-pad sensors (FBK and IHEP) irradiated at 1.5 and 2.5$\times$10$^{15}$ 1 MeV n$_{eq}/{cm}^2$.  }
\label{hgtdtime}
\end{figure}

\begin{figure}
\centering
\resizebox{0.5\textwidth}{!}{
  \includegraphics{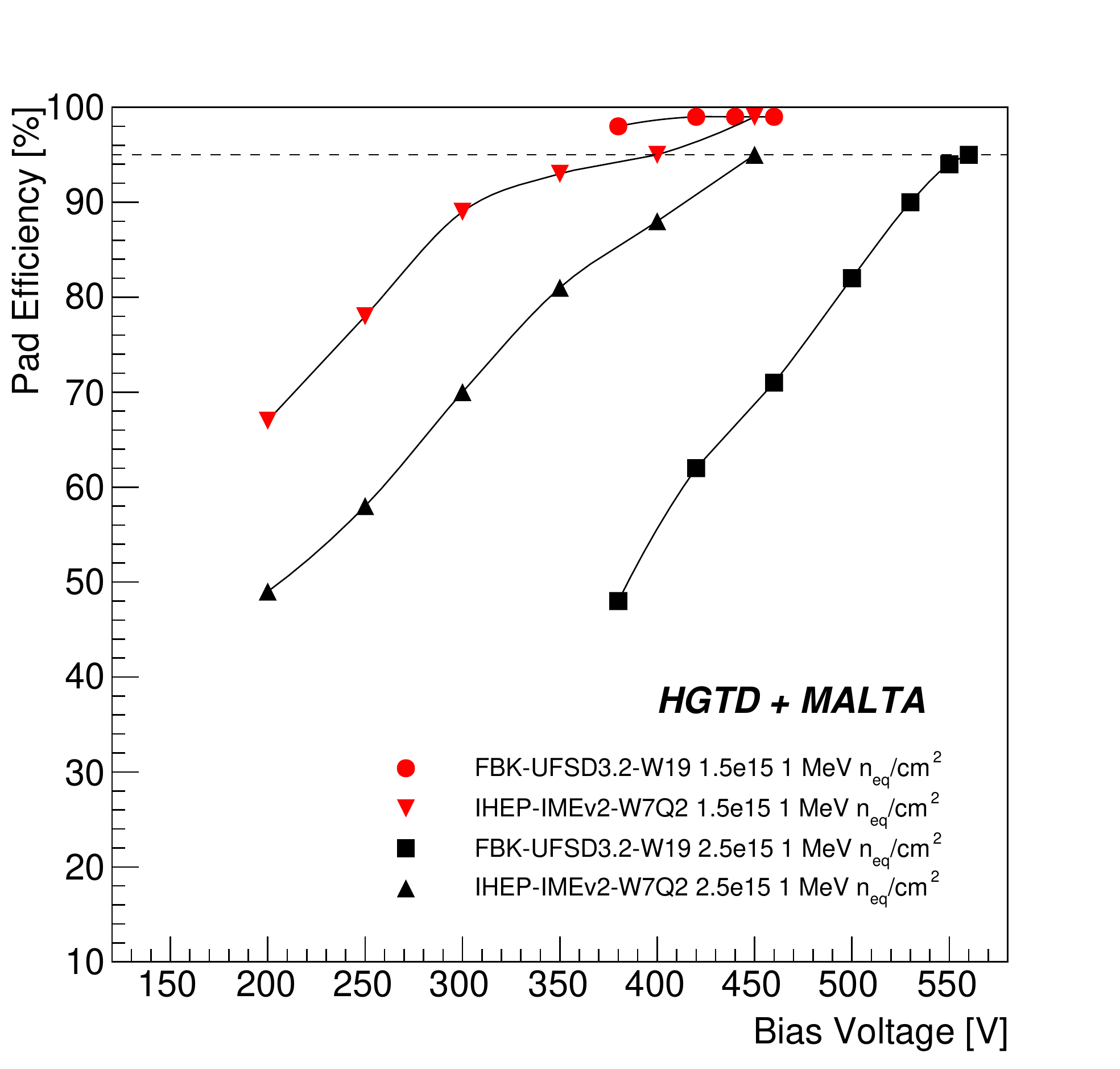}
}
\caption{Efficiency as a function of bias voltage for the HGTD test beam campaign in the MALTA telescope.The dashed lines corresponds to the minimum requirements for the future HGTD. The results are shown for different single-pad sensors (FBK and IHEP) irradiated at 1.5 and 2.5$\times$10$^{15}$ 1 MeV n$_{eq}/{cm}^2$.  }
\label{hgtdeff}
\end{figure}

\begin{figure}
\centering
\resizebox{0.5\textwidth}{!}{
  \includegraphics{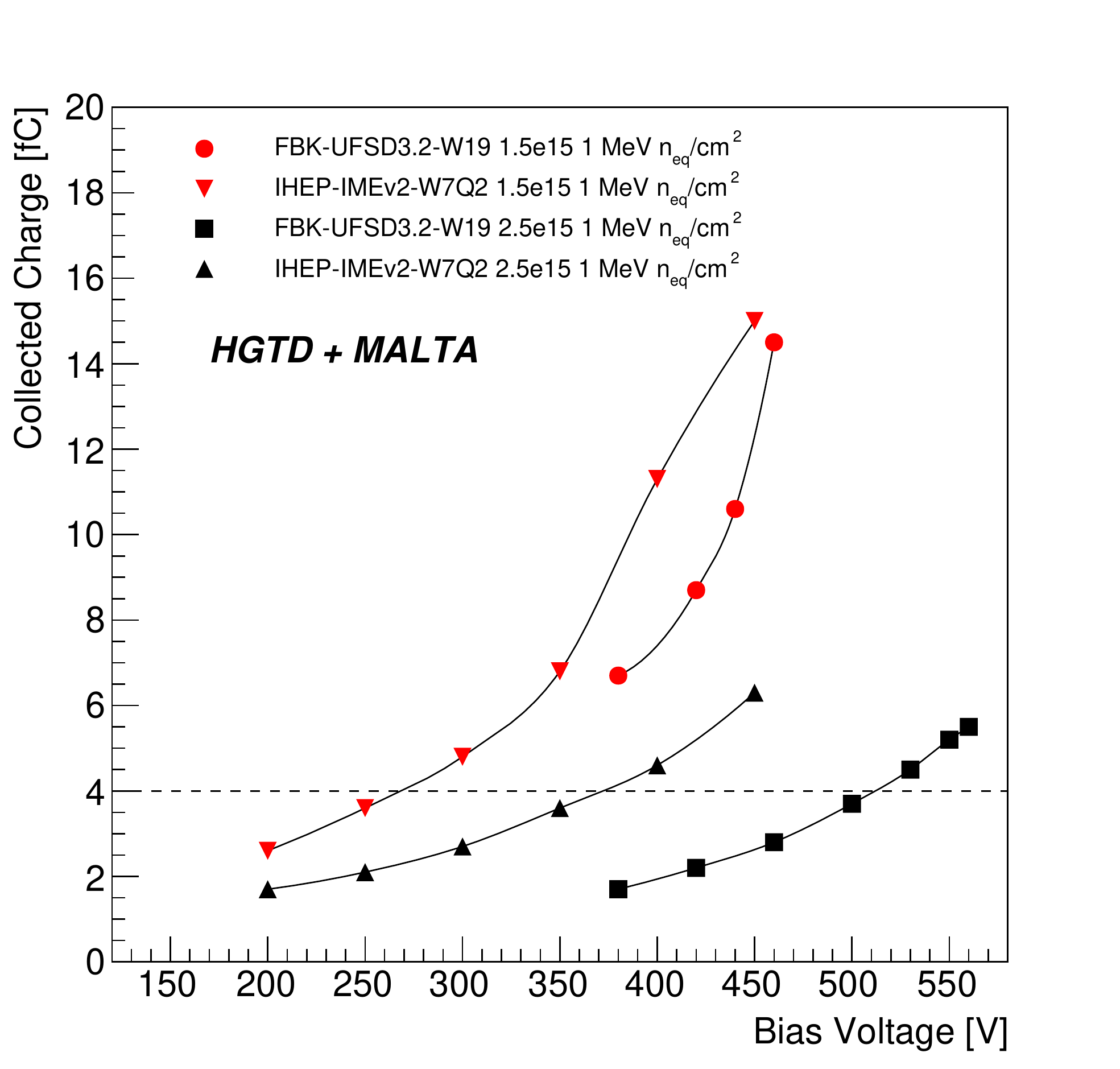}
}
\caption{Collected charge as a function of bias voltage for the HGTD test beam campaign in the MALTA telescope. The dashed lines corresponds to the minimum requirements for the future HGTD. The results are shown for different single-pad sensors (FBK and IHEP) irradiated at 1.5 and 2.5$\times$10$^{15}$ 1 MeV n$_{eq}/{cm}^2$.  }
\label{hgtdcharge}
\end{figure}

\noindent One of the key parameters that was measured during the test beam campaign was the time resolution of the LGAD sensors. The time resolution is extracted by subtracting the time of arrival of the DUT with the second LGAD installed for time reference. Figure \ref{hgtdtime} shows the time resolution for sensors irradiated at a fluence of 1.5 and 2.5$\times$10$^{15}$ 1 MeV n$_{eq}/{cm}^2$ as a function of the bias voltage. As can be observed in this plot, the time resolution improves with bias voltage and a time resolution of 40 ps for sensors FBK-2.5 (at 550 V) and 43 ps for IHEP-2.5 sensors (at 450 V) is achieved. Figure \ref{hgtdeff} shows that as the bias voltage increases, efficiencies up to approximately 99\% can be achieved for sensors irradiated to 1.5$\times$10$^{15}$ 1 MeV n$_{eq}/{cm}^2$. For sensors irradiated to higher fluences, efficiencies up to $\sim$95\% can be obtained. Finally, Figure \ref{hgtdcharge} shows that at a fluence of 1.5$\times$10$^{15}$ 1 MeV n$_{eq}/{cm}^2$, the collected charge is above the minimum required amount of 4 fC.  At higher fluence, the collected charge at the same bias voltage is less, though the HGTD requirements are still fulfilled at higher bias voltage. These results confirm the feasibility of an LGAD-based timing detector for HL-LHC. More extensive analysis and description of this test beam campaign can be found in Ref. \cite{lgads}.

\section{Conclusion}

Since 2021, the MALTA telescope has been extensively used for the characterization of R\&D prototypes during test beams in the H6 beamline of the North Area at CERN. The MALTA telescope is based on six monolithic MALTA tracking planes, which are fabricated in Tower 180 CMOS imaging technology. The MALTA telescope exploits the best qualities of the MALTA sensor: a full prototype (large area), high granularity, self-triggering, and good spatial and timing resolution. This allowed to build a telescope that offers a high degree of flexibility and versatility, and allows to test a wide variety of prototypes that target different requirements. The MALTA telescope provides a spatial resolution of $\sigma_{s} = 4.1\pm 0.2  ~\mu$m, based on the linear regression approach, and a track timing resolution of $\sigma_{t} = 2.1$ ns. Continuous detector R\&D activities will be a crucial ingredient in order to achieve ultimate performance at future colliders and therefore the continuous development and improvement of test beam telescopes is essential.

\begin{acknowledgement}

This project has received funding from the European Union’s Horizon 2020 Research and Innovation programme under Grant Agreement number 101004761 (AIDAinnova), and number 654168 (IJS, Ljubljana, Slovenia). Furthermore it has been supported by the Marie Sklodowska-Curie Innovative Training Network of the European Commission Horizon 2020 Programme under contract number 675587 (STREAM).

\end{acknowledgement}

\end{document}